\documentclass[preprint,prd,showpacs,amsmath,amssymb,amsthm,nofootinbib]{revtex4}
\usepackage[mathscr]{euscript}
\usepackage{bm}
\usepackage{graphicx}
\usepackage{subfigure}
\usepackage{overpic} 
\usepackage{multirow}  
\usepackage{floatrow}
\usepackage[colorlinks=true,linkcolor=red]{hyperref}
\newfloatcommand{capbtabbox}{table}[][\FBwidth]
\newcommand{\bea}{\begin{eqnarray}}
\newcommand{\eea}{\end{eqnarray}}
\newcommand{\beq}{\begin{equation}}
\newcommand{\eeq}{\end{equation}}

\def\/{\over}

\begin{document}

\title{Nonlinear Preheating with Nonminimally Coupled Scalar Fields in the Starobinsky Model }

\author{ Chengjie Fu$^{1}$,  Puxun Wu$^{1}$\footnote{Corresponding author: pxwu@hunnu.edu.cn} and Hongwei Yu$^{1}$\footnote{Corresponding author: hwyu@hunnu.edu.cn}  }
\affiliation{$^1$Department of Physics and Synergetic Innovation Center for Quantum Effects and Applications, Hunan Normal University, Changsha, Hunan 410081, China 
}

\begin{abstract}
We study the preheating after inflation in the Starobinsky model with a nonminimally coupled scalar field $\chi$. Using the lattice simulation, we analyze the rescattering between the $\chi$ particles and the inflaton condensate, and the backreaction effect of the scalar metric perturbations. We find that the rescattering is an efficient mechanism promoting the growth of the $\chi$ field variance. 
Meanwhile, copious inflaton particles can be knocked out of the inflaton condensate by rescattering. As a result, the inflaton field can become a non-negligible gravitational wave source, even comparable with the $\chi$ field in some parameter regions. For the scalar metric perturbations, which  are on the sub-Hubble scale in our analysis, our results show that they have negligible effects on the evolution of scalar fields and the production of gravitational waves in the model considered in present paper.

\end{abstract}

\pacs{98.80.Cq, 04.50.Kd, 05.70.Fh}

\maketitle
\section{Introduction}
\label{sec_in}
Inflation~\cite{Starobinsky,Inflation}, a phase of accelerated, quasiexponential expansion in the early Universe, is proposed to resolve the horizon, flatness and monopole problems, which plague the big bang standard cosmology. Meanwhile, the super-Hubble density perturbations  during inflation provide the seed for the formation of the large-scale structure in the Universe \cite{Structure}. In most models, inflation is driven by the  scalar field, called the inflaton, whose potential energy dominates over the kinetic energy. In contrast to these scalar field models, the Starobinsky model of inflation proposed in 1980 is characterized by a pure gravitational action of the form $R+\alpha R^2$ \cite{Starobinsky}, where $R$ is the Ricci scalar and $\alpha$ is a constant. After a conformal transformation, the Starobinsky model can be reformulated as a standard single-field slow-roll inflationary model \cite{1988Maeda}, which is included in a class of the $\alpha$-attractor E-models with the scalar field potential being $V(\phi)=\Lambda^4\left|1-\exp\left(-\frac{2\phi}{M}\right)\right|^{2n}$ \cite{2013Linde} and corresponds to the case of $n=1$ and $M\sim M_{p}$, where $M_p=2.4\times10^{18}\;{\rm GeV}$ is the reduced Planck mass.  For the Starobinsky model, assuming the number of \textit{e}-folds $N=60$, one can get the spectral index $n_s=1-\frac{2}{N}\simeq0.9667$ and the extremely low tensor-to-scalar ratio $r=\frac{12}{N^2}\simeq0.0033$ in the Einstein frame, which are in excellent agreement with the 2018 Planck data \cite{Planck2018}. The Starobinsky model is now regarded as one of the most popular inflationary models due to its simplicity and great consistency with observations.

The stage following inflation is called reheating \cite{reheating}, in  which the energy of the inflaton transfers into the thermal energy of elementary particles and the Universe reaches a radiation-dominated state that is necessary for a successful big bang nucleosynthesis. The elementary theory of reheating is based on the perturbation theory, but it has been recognized  that the first stage of reheating is a nonperturbative process called preheating \cite{preheating1, preheating2,2017Fu,2018Wang} in many inflationary models, as well as in bouncing cosmologies \cite{2011cai1,2011cai2}. The particle production of the daughter fields due to the parametric resonance during preheating is extremely efficient compared with the elementary theory of reheating. Furthermore, the preheating is an attractive phase with some interesting physical processes, such as the production of a significant spectrum of gravitational waves (GWs) \cite{2006Easther,2007Easther,2018Liu,2018Fu} and the primordial black holes \cite{2001Green,2006Suyama,2014Torres-Lomas}. During this phase, the homogeneous inflaton condensate pumps energy not only into other matter fields coupled to the inflaton field, but also into its own fluctuations by self-resonance. The simplest model of self-resonance is the $\lambda\phi^4$ model, whose nonlinear calculations of preheating were performed in Ref. \cite{1996Khlebnikov}, but this model is strongly disfavored by the Planck data due to its high tensor-to-scalar ratio. Recently, Lozanov and Amin \cite{2017Mustafa,2018Mustafa} found that the self-resonance can lead to interesting nonlinear effects on the formation of long- and short-lived localized field configurations (oscillons and transients) for some observationally favored models, for example, the $\alpha$-attractor T-models, the $\alpha$-attractor E-models, and monodromy-type potentials. However, in the Einstein frame of the Starobinsky model, the inflaton field is absent of self-resonance. Thus, coupling to other fields is necessary to achieve an effective preheating. In Ref. \cite{1999Tsujikawa1}, Tsujikawa \emph{et al}. proposed an extended Starobinsky model with a nonminimal coupling between the curvature scalar and a new scalar field $\chi$, whose Lagrangian reads  
\begin{align}\label{1}
\mathcal{L}=\sqrt{-g}\left[\frac{1}{2\kappa^2}\left(R+\frac{R^2}{6\mu^2}\right)-\frac{1}{2}\xi R \chi^2-\frac{1}{2}(\nabla\chi)^2-\frac{1}{2}m_{\chi}^2\chi^2\right]\;,
\end{align}   
where $\kappa^{-1}=M_p$, $\mu$ is fixed at $1.3\times10^{-5}M_p$ by the magnitude of the primordial density perturbations \cite{2007Faulkner}, $\xi$ is an arbitrary coupling parameter, and $m_{\chi}$ is the bare mass of the scalar field $\chi$. The above model is a phenomenological extension of the Starobinsky model, but its underlying physics can be advocated by some particular effects of quantum gravity. For example,  it was generically acknowledged that the $R^2$ term or even higher-order terms may arise naturally in the framework of the asymptotically safe gravity which can be applied to realize the inflationary phase at relatively low-energy scales 
  \cite{2012cai}. Later, the reheating process of the inflationary cosmology motivated by the asymptotically safe gravity has been examined in Ref. \cite{2013cai} at the perturbative level in the elementary theory of reheating. Basing on the Hartree approximation, the preheating process of the extended Starobinsky model given in Eq. (\ref{1}) has been studied in Ref.  \cite{1999Tsujikawa1} by making a conformal transformation [ $\hat g_{\mu\nu}=\Omega^2 g_{\mu\nu}$ and $\Omega^2=1-\xi\kappa^2\chi^2+R/(3\mu^2)$] and introducing a scalar field $\phi \equiv \sqrt{3/2}\kappa^{-1}\ln{\Omega^2}$ as the inflaton. It was found that the growth rate of the fluctuations of the $\chi$ field is improved greatly and the preheating is more efficient compared with that in  the chaotic inflation model with a nonminimally coupled $\chi$ field \cite{1998Bassett,1999Tsujikawa2}.
Recently, the dynamics and reheating processes of the modified Starobinsky models have been investigated in \cite{Lee18}. In addition, an inflationary model similar to Eq. (\ref{1}), called the mixed Higgs-$R^2$ inflationary model, has been proposed in \cite{Ema2017}, and its preheating process was discussed in \cite{He2019}.   

However, in the Hartree approximation, the rescattering between the inflaton condensate and $\chi$ particles and the backreaction effect of the metric perturbations cannot be taken into account~\cite{1999Tsujikawa1}. But, these  effects may have non-negligible impacts on preheating. For example, in the chaotic inflation model with $V(\phi,\chi)=\frac{1}{4}\lambda\phi^4+\frac{1}{2}g^2\phi^2\chi^2$, the rescattering between the $\chi$ particles and the inflaton condensate limits the growth of fluctuations of the $\chi$ field \cite{1997Khlebnikov2}, and the metric perturbations have a large impact on preheating, which greatly enhance the final abundance of the field variances \cite{2000Bassett}. Moreover, Bastero-Gil \textit{et al}. \cite{2008Bastero-Gil,2010Bastero-Gil} found that the amplified scalar metric perturbations do enhance the GW stochastic background  produced during preheating in a generic supersymmetric model of hybrid inflation. Thus, it would be interesting to perform fully nonlinear analyses of preheating for the model in Eq. (\ref{1}) to investigate the effects of the rescattering and the metric perturbations. 

In this paper, we will use the three-dimensional lattice simulation to investigate preheating including the scalar metric perturbations in the Starobinsky model with a nonminimally coupled scalar field. We organize our paper as follows: Sec. \ref{sec2} gives the basic equations and initial conditions of the field and metric variables. In Sec.~\ref{sec3}, we present the numerical results and analyze the evolution of scalar fields. The equation of state is studied  in Sec.~\ref{sec4} and the effect of the scalar metric perturbations is discussed in Sec.~\ref{sec5}. Finally, our conclusions and discussions are given in Sec.~\ref{sec6}.

\section{basics equations}
\label{sec2}
After a conformal transformation,  the system given in Eq.~(\ref{1}) can be expressed  to be the   Einstein frame one with the Lagrangian taking the form
\begin{align}\label{2}
\mathcal{L}=\sqrt{-\hat{g}}\left[\frac{1}{2\kappa^2}\hat R-\frac{1}{2}(\hat\nabla\phi)^2-\frac{1}{2}e^{-\sqrt{\frac{2}{3}}\kappa\phi}(\hat\nabla\chi)^2-V(\phi,\chi)\right]\;,
\end{align}
where $\phi$ is the introduced scalar field and  
\begin{align}\label{3}
V(\phi,\chi)=e^{-2\sqrt{\frac{2}{3}}\kappa\phi}\left[\frac{3\mu^2}{4\kappa^2}\left(e^{\sqrt{\frac{2}{3}}\kappa\phi}-1+\xi\kappa^2\chi^2\right)^2+\frac{1}{2}m_{\chi}^2\chi^2\right]\;.
\end{align}
Afterwards, since we work only in the Einstein frame, the caret will be dropped  in the following discussion. In the inflationary era, the $\chi$ field does not need to be taken into consideration and the $\phi$ field plays the role of an inflaton field. 

After the end of inflation, the $\phi$ field enters the coherent oscillation phase in which the $\phi$ field behaves as an inflaton with the quadratic potential around $\phi=0$. The parametric resonance of the $\chi$ particles occurs by the tachyonic instability, and the copious $\chi$ particles of small-momentum modes are excited. This is because the coupling of the $\chi$ field and inflaton field gives a tachyonic mass to the $\chi$ field, whose square of the effective mass is defined as
\begin{align}\label{4}
m^2_{\chi,\text{eff}}=\frac{d^2V}{d\chi^2}= e^{-2\sqrt{\frac{2}{3}}\kappa\phi}\left[3\mu^2\xi\left(e^{\sqrt{\frac{2}{3}}\kappa\phi}-1+3\xi\kappa^2\chi^2\right)+m_{\chi}^2\right]\;.
\end{align}
Although there is no $\phi$ resonance in this model, the produced $\chi$ particles knock inflaton quanta out of the condensate and into low-momentum modes. The growth of the inflaton fluctuations can be expected, and it is interesting to investigate the impacts of them on preheating. Before studying these processes, 
  let us give the basic equations of this model during preheating. In order to add the metric perturbations to the nonlinear calculations simultaneously, we use the Arnowitt-Deser-Misner metric \cite{1962ADM}, whose spacetime line element reads 
  \begin{align}\label{5}
ds^2&=g_{\mu\nu}dx^\mu dx^\nu \nonumber \\
&=-N^2dt^2+\gamma_{ij}(dx^i+N^idt)(dx^j+N^jdt)\;,
\end{align}
where $N$ is the lapse function, $N^i$ is the shift vector field, and $\gamma_{ij}$ is the spatial metric. The lapse and shift vector are gauge functions. Although the Newtonian gauge is usually optimal, it is very  difficult to investigate the nonlinear preheating under it, so we will work in the synchronous gauge with $N=0$ and $N^i=0$. Then we perform a conformal transformation of the spatial metric, $\tilde\gamma_{ij}=e^{-2\beta}\gamma_{ij}$, with $e^{2\beta}=\det(\gamma_{ij})^{1/3}$. The variable $\beta$ represents one scalar degree of freedom, and the spatial metric $\tilde\gamma_{ij}$ contains two tensor degrees of freedom, two vector degrees of freedom, and one traceless scalar degree of freedom. For simplicity, we neglect the perturbations of the spatial metric $\tilde\gamma_{ij}$ with $\tilde\gamma_{ij}=\delta_{ij}$ and only consider the effect of the scalar metric variable $\beta$. So, the spacetime metric (\ref{5}) can be rewritten as
\begin{align}\label{6}
ds^2=-dt^2+e^{2\beta}\delta_{ij}dx^idx^j\;.
\end{align}
The averaged scale factor is given by $a(t)\equiv\langle e^{3\beta}\rangle^{1/3}$ (where $\langle\cdots\rangle$ denotes the spatial average), and the Hubble parameter is defined to be $H(t)\equiv\dot a/a$. Thus,  one can obtain the following motion equations   of two scalar fields and the metric variable:
\begin{align}\label{7}
\ddot\phi+3\dot\beta\dot\phi-e^{-2\beta}\nabla^2\phi-e^{-2\beta}\partial^k\beta\partial_k\phi+\frac{\kappa}{\sqrt{6}}e^{-\sqrt{\frac{2}{3}}\kappa\phi}\left(\dot\chi^2-e^{-2\beta}\partial^k\chi\partial_k\chi\right)+\frac{dV}{d\phi}=0\;,
\end{align}  
\begin{align}\label{8}
\ddot\chi+3\dot\beta\dot\chi-e^{-2\beta}\nabla^2\chi-e^{-2\beta}\partial^k\beta\partial_k\chi-\sqrt{\frac{2}{3}}\kappa\left(\dot\phi\dot\chi-e^{-2\beta}\partial^k\chi\partial_k\phi\right)+e^{\sqrt{\frac{2}{3}}\kappa\phi}\frac{dV}{d\chi}=0\;,
\end{align}
\begin{align}\label{9}
\ddot\beta+\dot\beta^2=-\frac{\kappa^2}{6}(\rho+3p)\;,
\end{align} 
where the total energy density and pressure are defined as
\begin{align}\label{10}
\rho=\frac{1}{2}\left(\dot\phi^2+e^{-\sqrt{\frac{2}{3}}\kappa\phi}\dot\chi^2\right)+\frac{1}{2}e^{-2\beta}\left((\partial\phi)^2+e^{-\sqrt{\frac{2}{3}}\kappa\phi}(\partial\chi)^2\right)+V(\phi,\chi)\;,
\end{align} 
\begin{align}\label{11}
p=\frac{1}{2}\left(\dot\phi^2+e^{-\sqrt{\frac{2}{3}}\kappa\phi}\dot\chi^2\right)-\frac{1}{6}e^{-2\beta}\left((\partial\phi)^2+e^{-\sqrt{\frac{2}{3}}\kappa\phi}(\partial\chi)^2\right)-V(\phi,\chi)\;.
\end{align} 
In addition, from the Hamiltonian constraint equation, one can obtain
\begin{align}\label{12}
3\dot\beta^2-2e^{-2\beta}\nabla^2\beta-e^{-2\beta}\partial^k\beta\partial_k\beta=\kappa^2\bigg[\frac{1}{2}\left(\dot\phi^2+e^{-\sqrt{\frac{2}{3}}\kappa\phi}\dot\chi^2\right) + \nonumber \\
\frac{1}{2}e^{-2\beta}\left(\partial^k\phi\partial_k\phi+e^{-\sqrt{\frac{2}{3}}\kappa\phi}\partial^k\chi\partial_k\chi\right)+V(\phi,\chi)\bigg]\;.
\end{align} 
 
The initial conditions of the preheating are determined by the dynamics of inflation.  The inflation takes place when the value of the $\phi$ field is larger than $M_p$, and ends when the Hubble slow-roll parameter $\epsilon\equiv -\dot H/H^2$ is equal to unity, at which point the values of the inflaton and its derivative are $\phi_e\simeq0.97M_{p}$ and $\dot\phi_e\simeq-3.75\times10^{-6}M_p^2$, respectively. It is reasonable to  treat the end of inflation as the onset of preheating. Thus,  the initial preheating values of the inflaton and its derivative can be set as $\phi_i=\phi_e$ and $\dot\phi_i=\dot\phi_e$. Therefore, the energy scale at the beginning of the preheating  is about $(6\times10^{15}\;{\rm GeV})^4$, which is less than the inflationary scale. Since the matter field $\chi$ is negligible at the beginning of preheating,  the $\chi$ field and its derivative are initialized as zero.  The fluctuations of two scalar fields and their derivatives are initialized by quantum vacuum fluctuations. For convenience, we initialize the scale factor as $a_i=1$, which means that the scalar metric variable can be initialized as $\beta_i=0$. The initial value of $\dot\beta$ can be obtained  from Eq.~(\ref{12}).

\section{nonlinear preheating process}
\label{sec3}
Using a modified version of the publicly available Fortran package HLattice \cite{2011Huang}, where the fourth-order Runge-Kutta integrator takes the place of the symplectic one, we perform numerical lattice simulations with $128^3$ points to investigate the evolutions of the field and metric variables during preheating in the model in Eq. (\ref{2}).
In our analysis, we choose the lattice length of side $L$ to satisfy $LH<  2\pi$, which means that all field modes are within the horizon at the beginning of the simulation. 
   In the following sections, for convenience  we consider the cases of $m_\chi=0$ and $m_\chi>0$,  respectively.

\subsection{Massless $\chi$ particle case}\label{3_A}
\begin{figure}
\centering
\includegraphics[width=0.8\textwidth ]{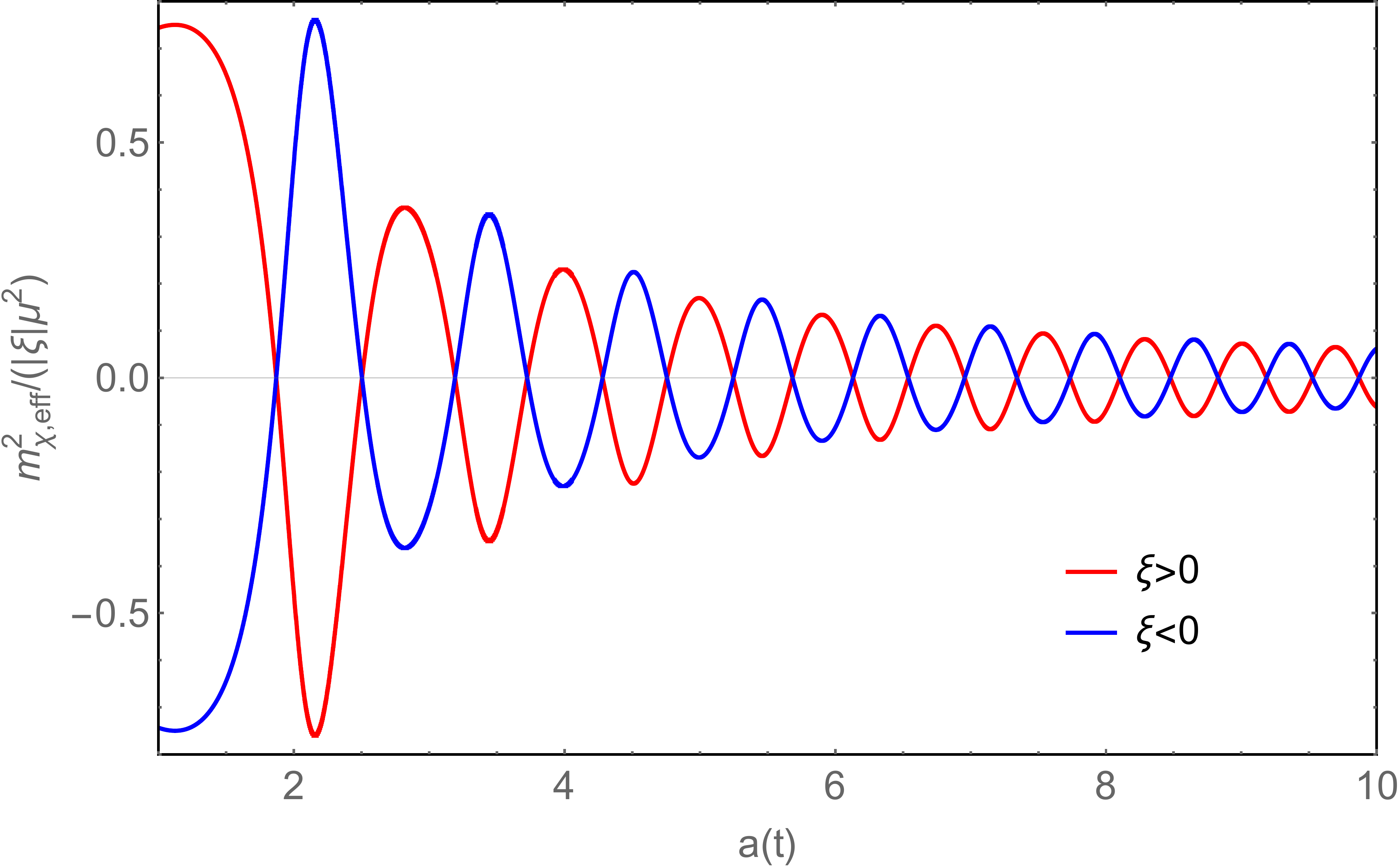}
\caption{\label{fig1} The evolution of $m^2_{\chi,\text{eff}}/(|\xi|\mu^2)$ as a function of $a(t)$ for  the case of $m_\chi=0$.}
\end{figure}

For the case of $m_\chi=0$, first we neglect the backreaction effect of the $\chi$ particles, and then Eq. (\ref{4}) becomes
\begin{align}\label{13}
m^2_{\chi,\text{eff}}\simeq 3\xi e^{-2\sqrt{\frac{2}{3}}\kappa\phi}\left(e^{\sqrt{\frac{2}{3}}\kappa\phi} -1 \right)\mu^2\;.
\end{align}
The evolution of $m^2_{\chi,\text{eff}}/(|\xi|\mu^2)$ as a function of the scale factor $a(t)$ is shown in Fig. \ref{fig1}. One can see that $m^2_{\chi,\text{eff}}$ oscillates around zero, which causes the parametric resonance. When $m^2_{\chi,\text{eff}}<0$, all modes with $k^2/a^2<|m^2_{\chi,\text{eff}}|$ experience exponential growth. The resonance intensity and the width of the resonance band are positively related to the amplitude of $m^2_{\chi,\text{eff}}$ that is determined  by the value of $|\xi|$. Unlike the case of $\xi>0$, the initial effective mass of the $\chi$ field is tachyonic when $\xi<0$. This characteristic makes the parametric resonance in the case of $\xi<0$ more efficient than that in the case of $\xi>0$ for the same $|\xi|$.

\begin{figure}
\centering
\includegraphics[width=0.8\textwidth ]{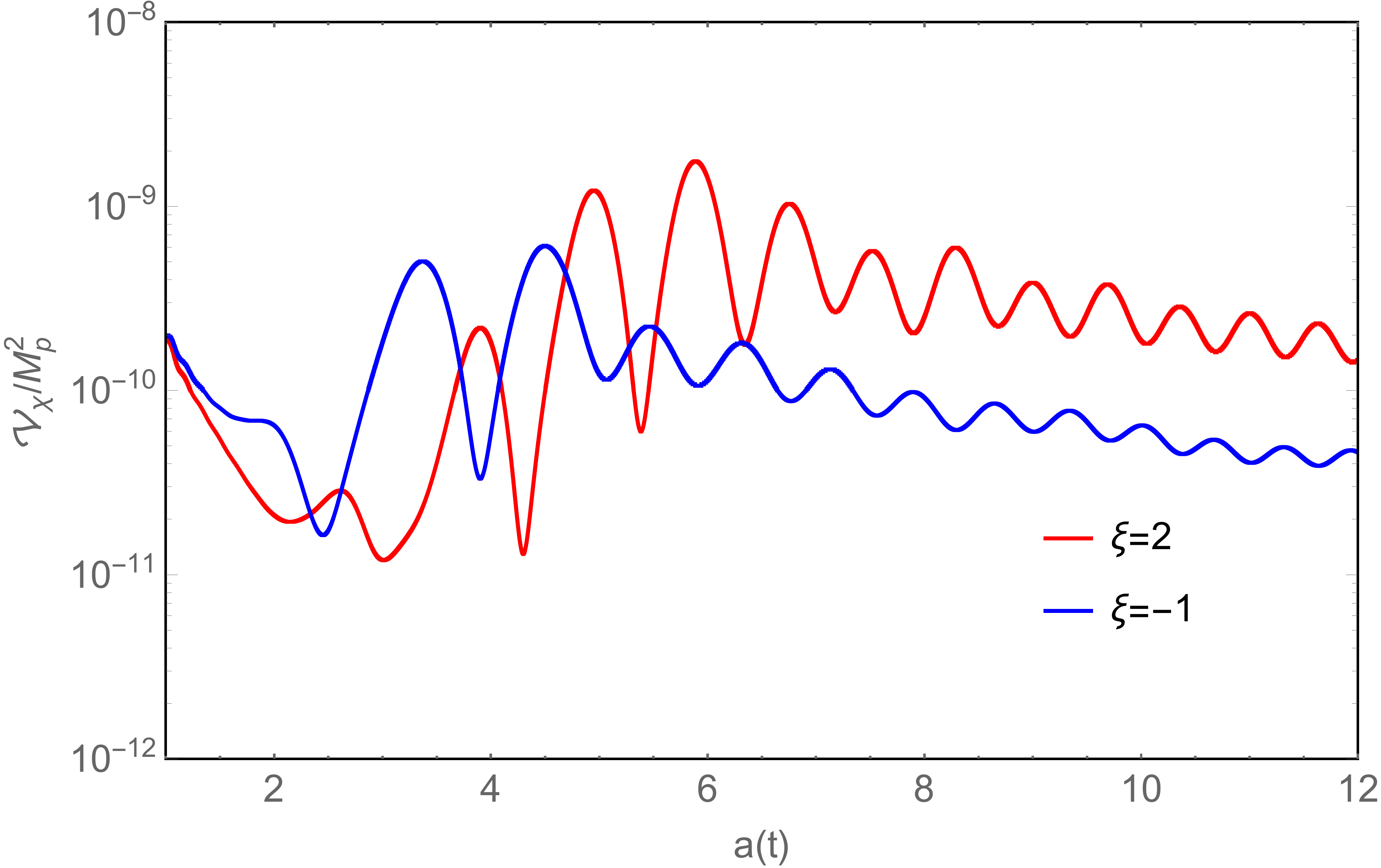}
\caption{\label{fig2} The variance of the $\chi$ field versus $a(t)$ for $m_\chi=0$.}
\end{figure}

Since the resonance intensity becomes stronger and stronger with the increase of $|\xi|$, for the case when $|\xi|$ is very small, i.e. $-1<\xi<2$, the resonance efficiency is not enough to fight the expansion of the Universe, which can also be found in Fig.~\ref{fig2}, where  the evolutions of the $\chi$ field variance $\mathcal{V}_\chi\equiv\langle \chi^2\rangle-\langle \chi\rangle^2$ as a function of $a(t)$ are plotted for $\xi=2$ (red line) and $\xi=-1$ (blue line). One can see that  the maximum of the $\chi$ field variance is almost the same as  its initial value in both cases, which means that the former will be less than the  latter when $-1<\xi<2$.  
  Apparently, both evolutions of  $\mathcal{V}_\chi$ 
  decrease initially due to the expansion of the Universe, but the blue line decays more slowly than the red one. This is because the resonance has happened before the inflaton field enters the coherent oscillation for the $\xi<0$ case.

Figure \ref{3} shows the semilog plot of the maximum of the $\chi$ field variance ($\mathcal{V}_{\chi,\text{max}}$) as a function of $\xi$. Obviously, $\mathcal{V}_{\chi,\text{max}}$ when $\xi<0$ is always greater than the value with $\xi>0$ for the same $|\xi|$.  The maximum of $\mathcal{V}_{\chi,\text{max}}$ is $1.99\times10^{-2}M_p^2$, which occurs at $\xi\simeq-5$. Furthermore, it is easy to see that $\mathcal{V}_{\chi,\text{max}}$ decreases as $1/|\xi|$ when $\xi\gtrsim 80$ and $\xi\lesssim-40$, which is consistent with the conclusion obtained with the Hartree approximation~\cite{1999Tsujikawa1}. In Tables \ref{table1} and \ref{table2}, we give a comparison for the $\mathcal{V}_{\chi,\text{max}}$ values obtained from  the lattice simulation and the Hartree approximation,  respectively, for several different values of $\xi$. We find that the growth of the $\chi$ fluctuations is enhanced after considering the nonlinear effects. To analyze clearly the evolutions of the inflaton field and the massless $\chi$ field during preheating, we now separate our discussion into two cases:  $\xi\geq0$ and $\xi<0$.

\begin{figure}
\centering
\includegraphics[width=0.8\textwidth ]{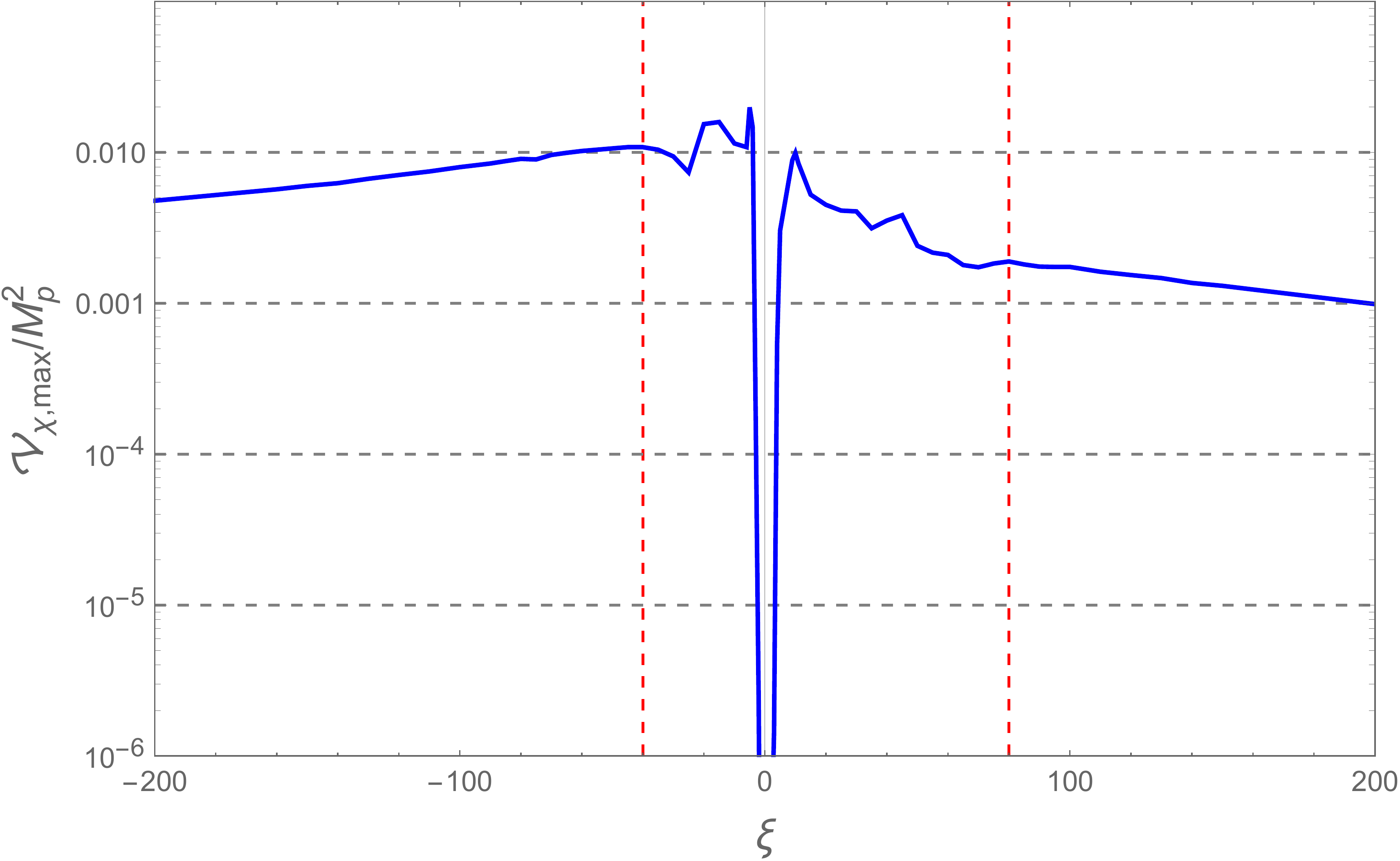}
\caption{\label{fig3} The maximum of the $\chi$ field variance as a function of $\xi$ for $m_\chi=0$ .}
\end{figure}

\begin{table*} 
 \begin{floatrow} 
 \capbtabbox{
   \begin{tabular}{ccc}
  \hline
  \hline
   \multirow{2}{*}{$\xi\qquad$}&\multicolumn{2}{c}{$\mathcal{V}_{\chi,\text{max}}/M_p^2$}\\
   \cline{2-3}
   & $\qquad$Lattice$\qquad$ & $\qquad$Hartree$\qquad$ \\  
  \hline
  \hline
  3$\qquad$  & $1.12\times10^{-6}$  &    $5.55\times10^{-7}$     \\
  10$\qquad$  & $9.90\times10^{-3}$  & $2.50\times10^{-3}$    \\
  30$\qquad$ &$4.06\times10^{-3}$   & $1.89\times10^{-3}$   \\
  50$\qquad$ &$2.41\times10^{-3}$   & $7.29\times10^{-4}$    \\
  100$\qquad$ &$1.74\times10^{-3}$   & $6.31\times10^{-4}$   \\
  \hline
  \hline
  \end{tabular}}
  {\caption{A comparison of the maximum of the $\chi$ field variance obtained in the lattice simulation and in the Hartree approximation with different values of  $\xi$ for $m_\chi=0$.}\label{table1}}
  \capbtabbox{
  \begin{tabular}{ccc}
  \hline
  \hline
   \multirow{2}{*}{$\xi\qquad$}&\multicolumn{2}{c}{$\mathcal{V}_{\chi,\text{max}}/M_p^2$}\\
   \cline{2-3}
   & $\qquad$Lattice$\qquad$ & $\qquad$Hartree$\qquad$ \\  
  \hline
  \hline
  $-3$ $\qquad$  &  $4.06\times10^{-3}$ &   $2.19\times10^{-3}$      \\
  $-10$ $\qquad$  & $1.14\times10^{-2}$  & $4.47\times10^{-3}$    \\
  $-30$ $\qquad$ &$9.39\times10^{-3}$   & $7.01\times10^{-3}$   \\
  $-50$ $\qquad$ &$1.06\times10^{-2}$   & $7.09\times10^{-3}$    \\
  $-100$ $\qquad$ &$7.97\times10^{-3}$   & $3.07\times10^{-3}$   \\
  \hline
  \hline
  \end{tabular}}
  {\caption{A comparison of the maximum of the $\chi$ field variance obtained in the lattice simulation and in the Hartree approximation with different values of $\xi$ for $m_\chi=0$.}\label{table2}}
 \end{floatrow} 
\end{table*}

\subsubsection{$\xi\geq0$}\label{3_A_1} 
\begin{figure}
\centering
\subfigure{\label{fig4a}}{\includegraphics[width=0.48\textwidth ]{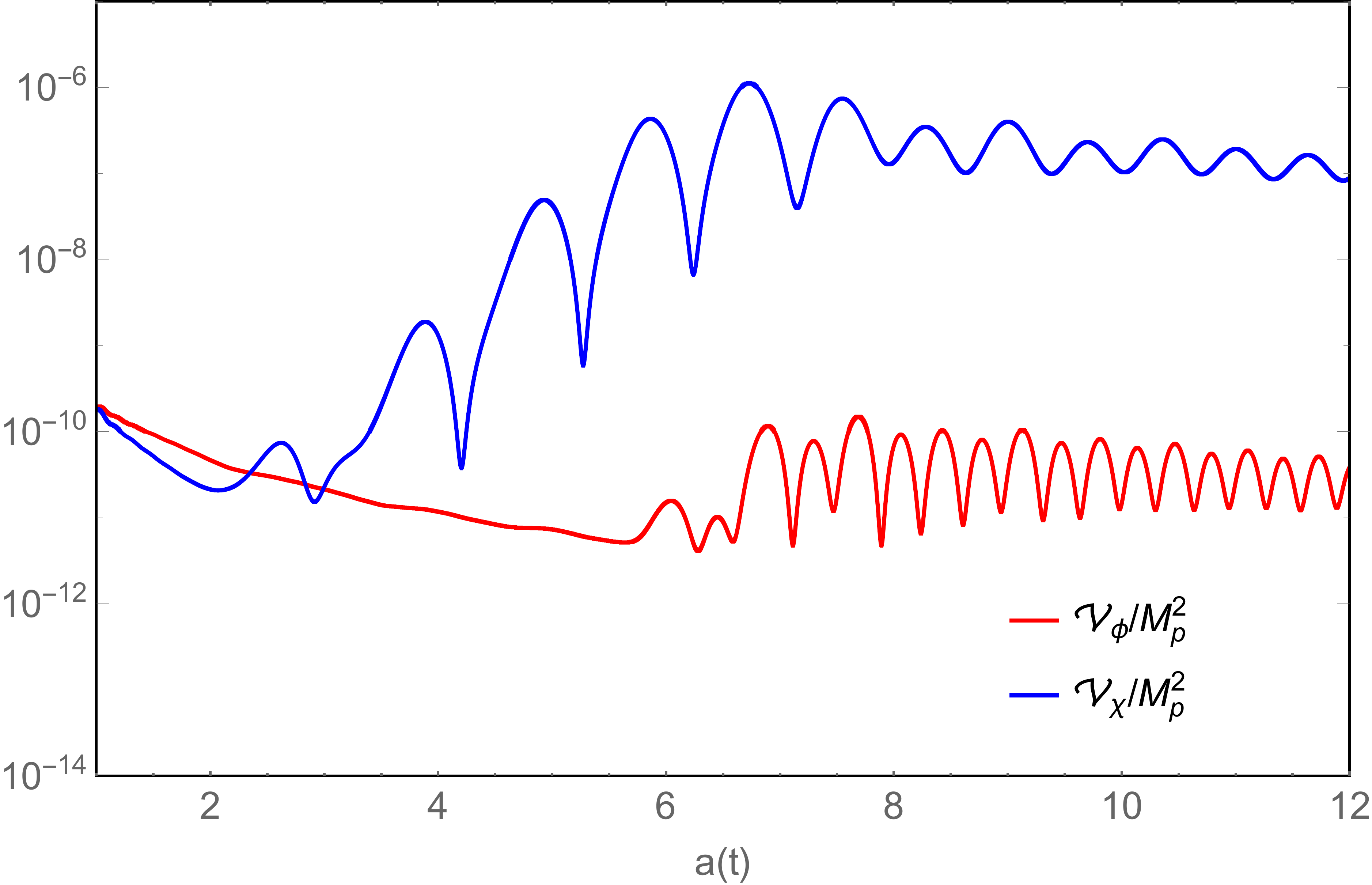}}
\subfigure{\label{fig4b}}{\includegraphics[width=0.48\textwidth ]{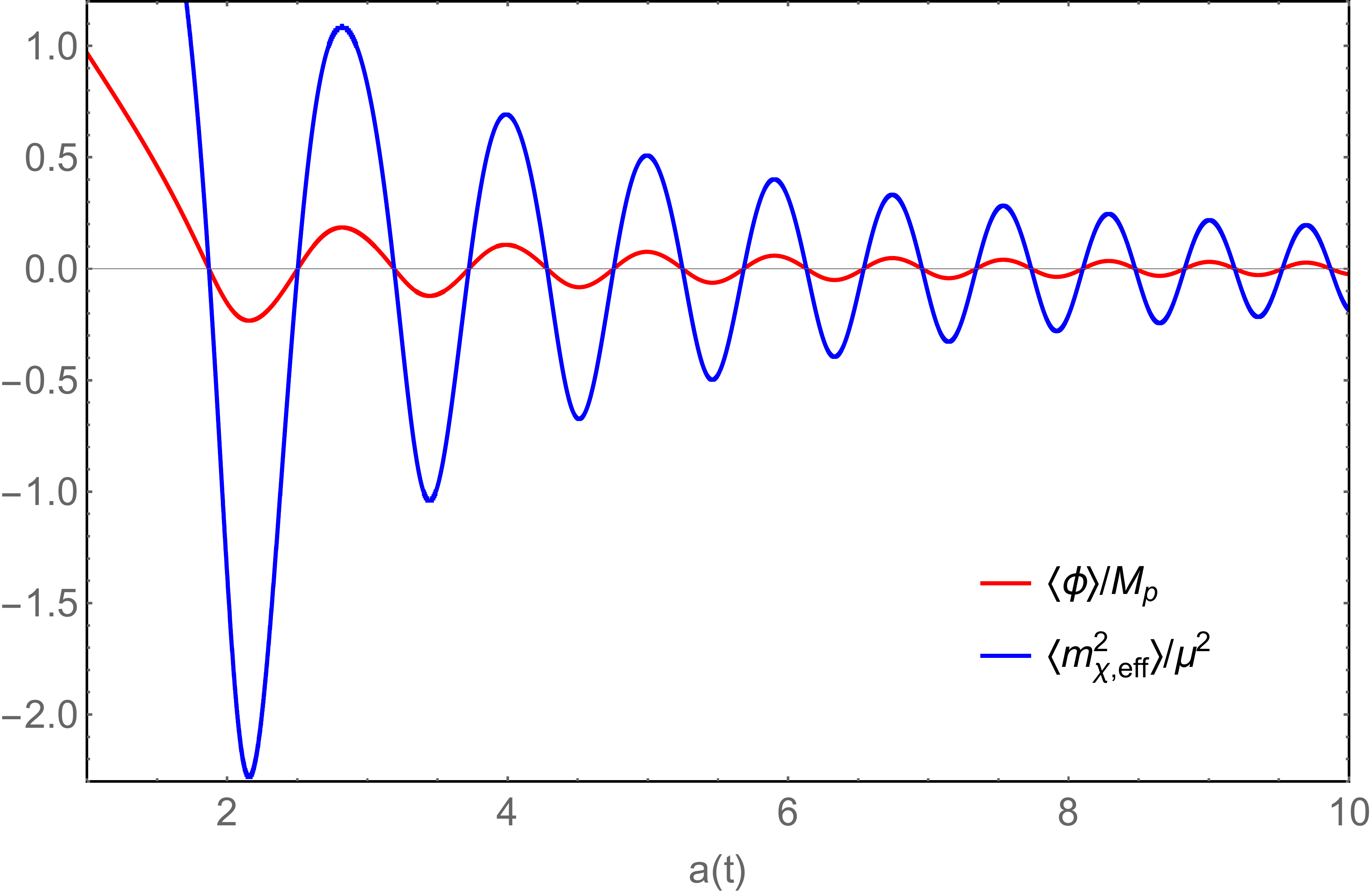}}
\caption{\label{fig4} (a) \textit{Left-hand plot}: The variances of the $\phi$ field (red line) and the $\chi$ field (blue line) versus $a(t)$ for $\xi=3$ and $m_\chi=0$. (b) \textit{Right-hand plot}: The spatial averages of the inflaton field $\langle\phi\rangle$ (red line) and $\langle m^2_{\chi,\text{eff}}\rangle$ (blue line) versus $a(t)$ for $\xi=3$ and $m_\chi=0$.}
\end{figure}

In Fig. \ref{fig4a}, we plot the evolutions of the variance of the inflaton field $\mathcal{V}_{\phi}\equiv\langle \phi^2\rangle-\langle \phi\rangle^2$ and the $\chi$ field variance $\mathcal{V}_{\chi}$ as a function of $a(t)$ for $\xi=3$. The variance of the inflaton field barely increases and is far less than that of the $\chi$ field. This is because few inflaton particles are knocked out of the condensate due to the poor abundance of $\mathcal{V}_{\chi}$. Thus,  the homogeneous part of the inflaton field $\langle\phi\rangle$ maintains coherent oscillation, and the spatial average of Eq. (\ref{4}) ($\langle m^2_{\chi,\text{eff}}\rangle$) also oscillates sustainedly around zero [see Fig. \ref{fig4b}]. Although the tachyonic mass of the $\chi$ particle always exists, the amplitude of $\langle m^2_{\chi,\text{eff}}\rangle$, associated with the amplitude of the oscillating inflaton field, decays with the expansion of the Universe. This effect makes the resonance become weak and eventually shuts off the growth of $\mathcal{V}_{\chi}$. 

\begin{figure}
\centering
\includegraphics[width=0.8\textwidth ]{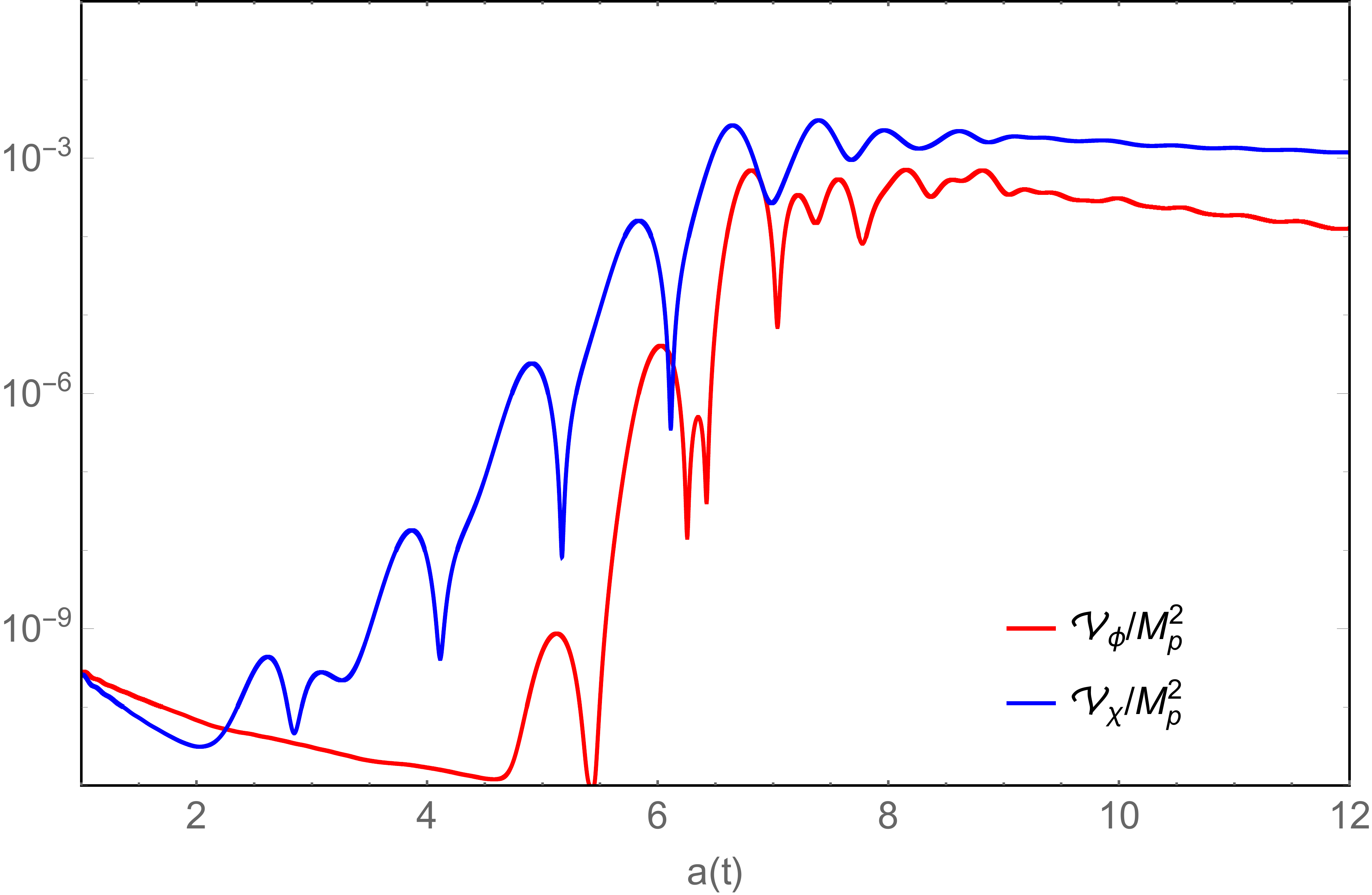}
\caption{\label{fig5} The variances of the $\phi$ field (red) and the $\chi$ field (blue) versus $a(t)$ for $\xi=5$ and $m_\chi=0$.} 
\end{figure}

\begin{figure}
\centering
\subfigure{\label{fig6a}}{\includegraphics[width=0.48\textwidth ]{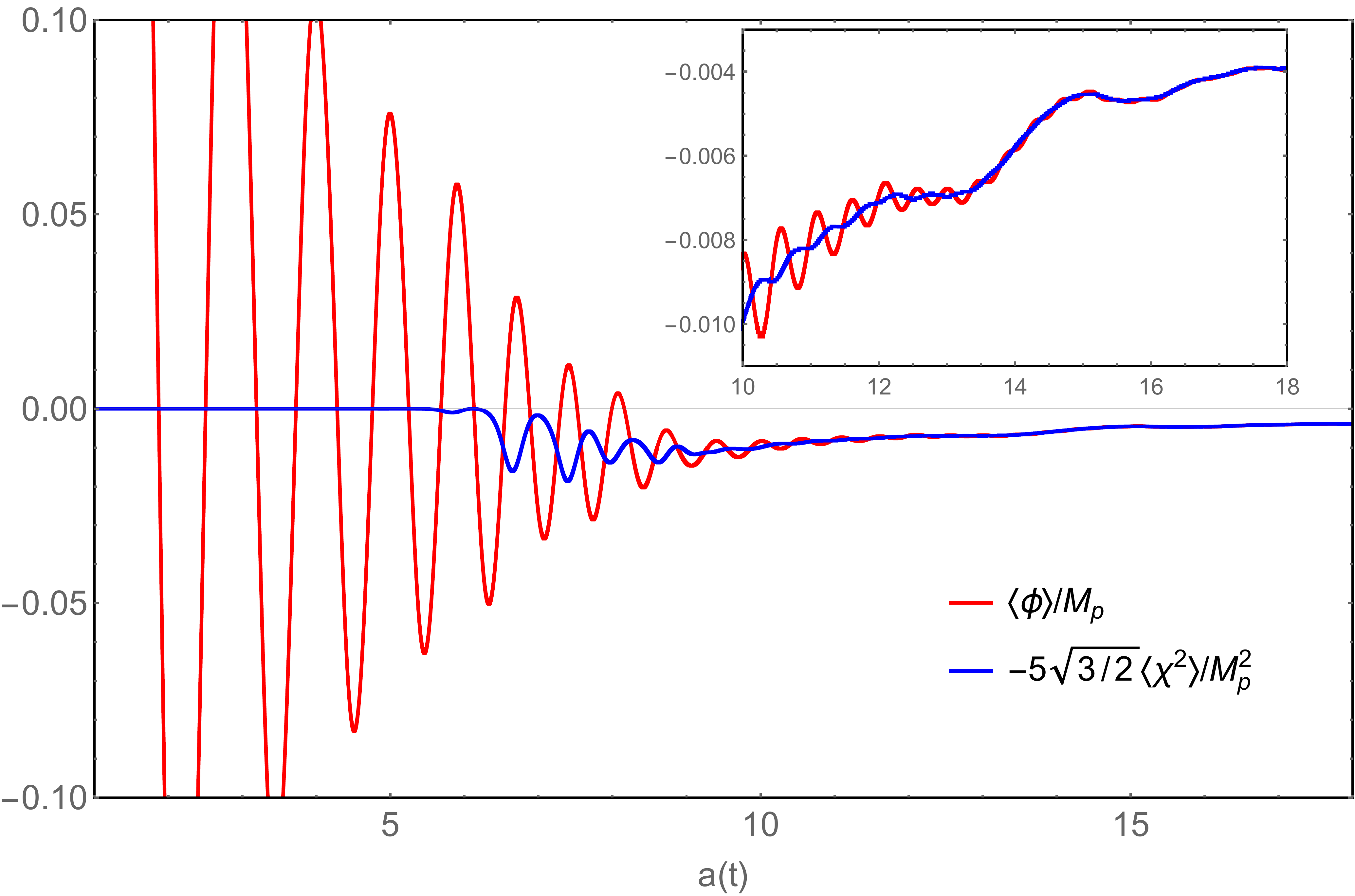}}
\subfigure{\label{fig6b}}{\includegraphics[width=0.48\textwidth ]{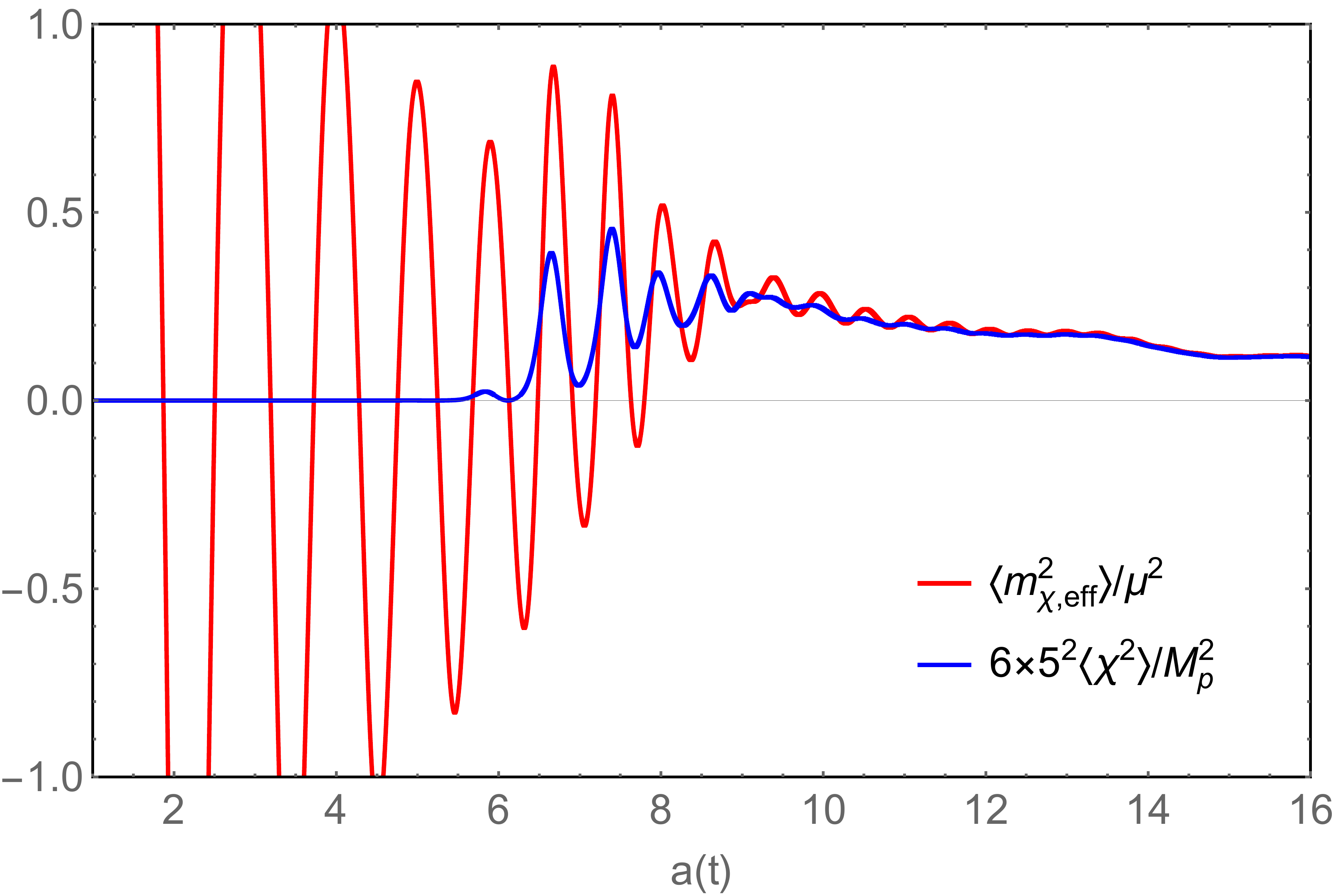}}
\caption{\label{fig6} (a) \textit{Left-hand plot}: The evolutions of $\langle\phi\rangle$ (red line) and $-5\sqrt{\frac{3}{2}}\langle\chi^2\rangle$ (blue line) as functions of $a(t)$ for $\xi=5$ and $m_\chi=0$. (b) \textit{Right-hand plot}: The evolutions of $\langle m^2_{\chi,\text{eff}}\rangle$ (red line) and $6\times5^2\langle\chi^2\rangle$ (blue line) as functions of $a(t)$ for $\xi=5$ and $m_\chi=0$.}
\end{figure}

Figure \ref{fig5} gives the evolutions of the variances of the $\phi$ and $\chi$ fields as a function of $a(t)$ for $\xi=5$. In this case, the stronger resonance makes the maximum of $\mathcal{V}_{\chi}$ far greater than that in the case of $\xi=3$. Since the production of more $\chi$ particles leads to stronger backreaction, the variance of the inflaton field increases significantly, although it is still less than the $\chi$ field variance.  The evolutions of $\langle\phi\rangle$ and $\langle m^2_{\chi,\text{eff}}\rangle$ are shown in Figs. \ref{fig6a} and \ref{fig6b}, respectively. From these, we find  that the energy transfer from the inflaton field to the $\chi$ field accelerates the decay of the amplitude of $\langle \phi\rangle$ and promotes the decay of the amplitude of $\langle m^2_{\chi,\text{eff}}\rangle$. Figure \ref{fig6a} indicates that $\langle\phi\rangle$ oscillates around   $\langle\phi\rangle=-\xi\sqrt{\frac{3}{2}}\kappa\langle\chi^2\rangle$ rather than $\langle\phi\rangle=0$. This property can be found in Eq.~(\ref{3}). 
   With the decay of the energy of the inflaton condensate,  $\langle\phi\rangle$ will finally stabilize at $\langle\phi\rangle=-\xi\sqrt{\frac{3}{2}}\kappa\langle\chi^2\rangle$.  Since $\langle m^2_{\chi,\text{eff}}\rangle$ will always be larger than zero and finally  stabilize at $\langle m^2_{\chi,\text{eff}}\rangle=6\xi^2\kappa^2\langle\chi^2\rangle\mu^2$, the tachyonic mass of the $\chi$ field will disappear completely and thus the parametric resonance will be shut off correspondingly. The ending time of parametric resonance is determined by both  the expansion of the Universe and the backreaction effect of the $\chi$ particles produced during preheating. 
   The case of $\xi\simeq5$ is the critical one, where the expansion of the Universe and the backreaction effect have  about equivalent contributions to  stopping the exponential growth of  $\mathcal{V}_{\chi}$. When $\xi\gtrsim 5$, the backreaction effect becomes important and increasingly dominant with the increase of $\xi$. 

\begin{figure}
\centering
\includegraphics[width=0.8\textwidth ]{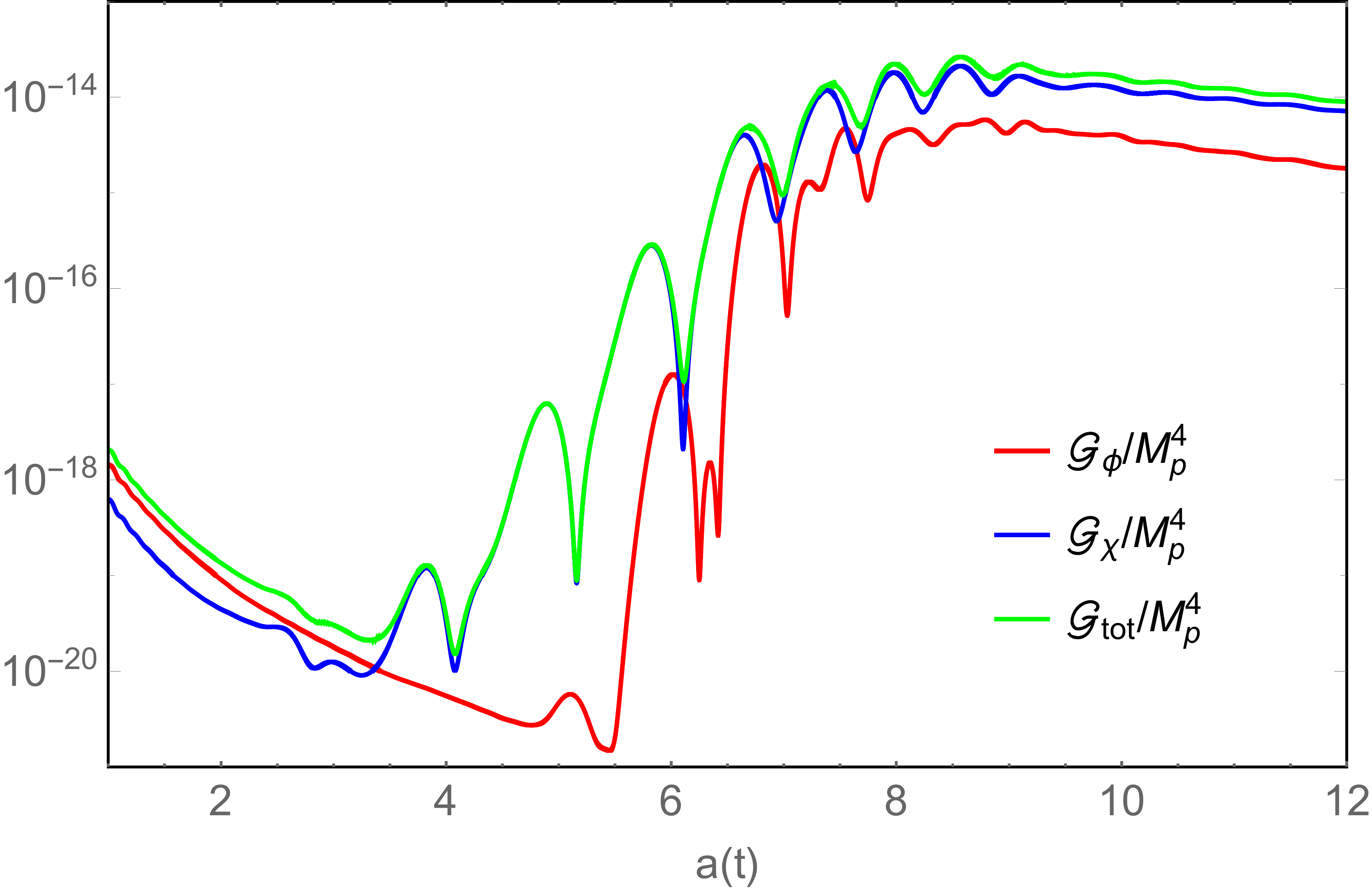}
\caption{\label{fig7} The evolutions of the average gradient energy density of the $\phi$ field (red line), the $\chi$ field (blue line), and their sum  (green line) with  $a(t)$ for $\xi=5$ and $m_\chi=0$.} 
\end{figure}

\begin{figure}
\centering
\subfigure{\label{fig8a}}{\includegraphics[width=0.48\textwidth ]{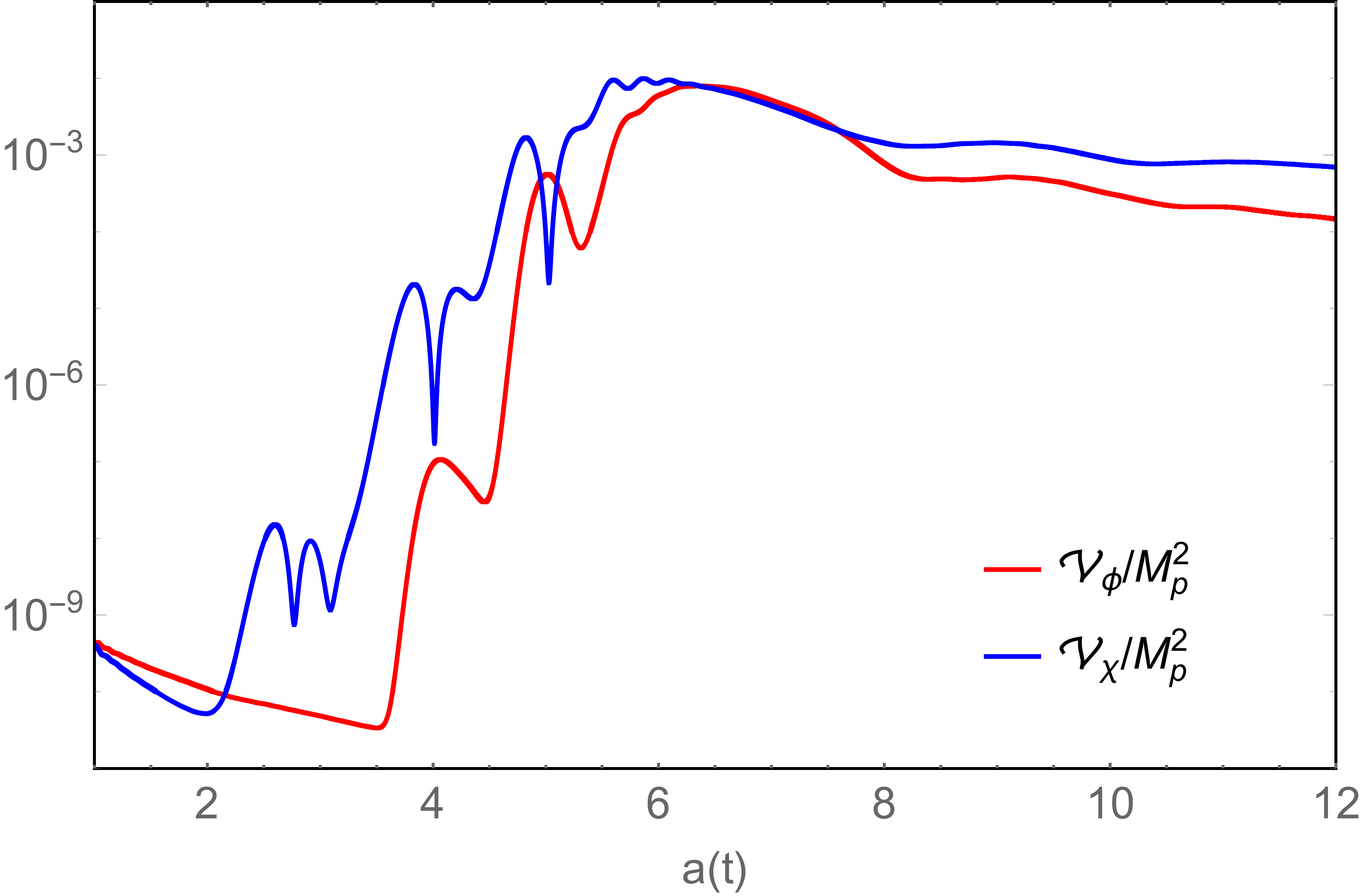}}
\subfigure{\label{fig8b}}{\includegraphics[width=0.48\textwidth ]{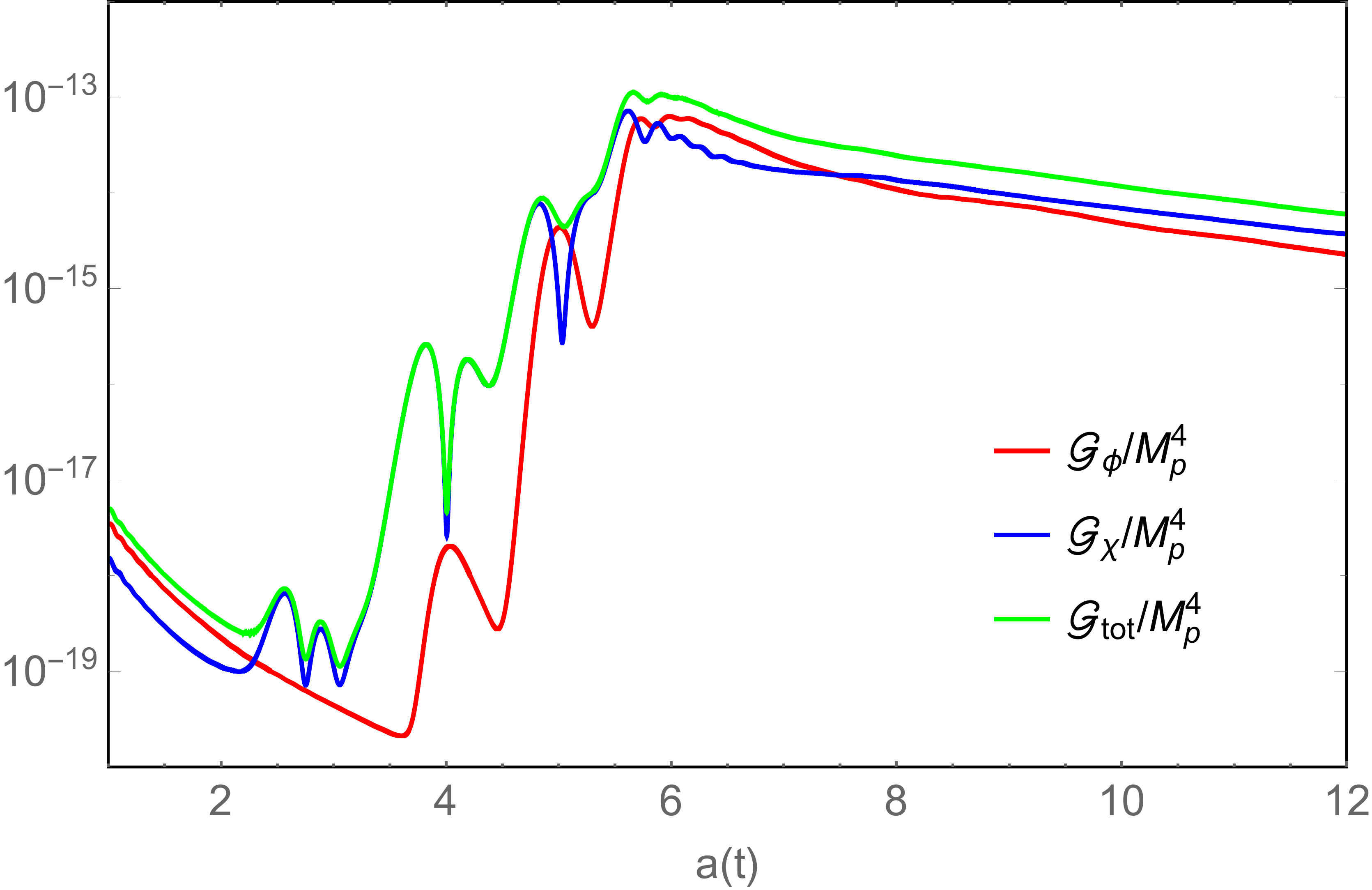}}
\caption{\label{fig8} (a) \textit{Left-hand plot}: The variances of the $\phi$ field (red) and the $\chi$ field (blue) versus $a(t)$ for $\xi=10$ and $m_\chi=0$. (b) \textit{Right-hand plot}: The evolutions of the average gradient energy density of the $\phi$ field (red line), the $\chi$ field (blue line), and their sum (green line) with  $a(t)$ for $\xi=10$ and $m_\chi=0$.}
\end{figure}

The production of a large number of the $\chi$ particles with nonzero modes will result in the matter distribution of the $\chi$ field having large density inhomogeneities in the position space. This means that the GWs can be sourced by the $\chi$ field, or more specifically, by its gradient. Although the fluctuations of the inflaton field are absent of the resonance, the copious inflaton particles can still be knocked out of the inflaton condensate by the $\chi$ particles created by the resonance. Thus, the inflaton field can also become an effective GW source. Figure \ref{fig7} shows the evolutions of the average gradient energy density of the $\phi$ field $\mathcal{G}_\phi\equiv \langle e^{-2\beta}(\partial\phi)^2\rangle/2$, the average gradient energy density of the $\chi$ field $\mathcal{G}_\chi\equiv \langle e^{-\sqrt{\frac{2}{3}}\kappa\phi}e^{-2\beta}(\partial\chi)^2\rangle/2$, and the average total gradient energy density $\mathcal{G}_{tot}\equiv \mathcal{G}_\phi+\mathcal{G}_\chi$ as a function of $a(t)$ for $\xi=5$. It is obvious that when $\mathcal{G}_{tot}$ reaches its maximum, the gradient of the inflaton field accounts for a small proportion of the total gradient, which means that the inflaton field excites few GWs compared with the $\chi$ field. If the parametric resonance can be further enhanced, more inflaton particles are knocked out, and the contribution of the inflaton field to GWs can be increased. For example, as shown in Fig. \ref{fig8a}, when  $\xi=10$, after the end of the growth of $\mathcal{V}_{\chi}$, the variance of the inflaton field is almost equal to that of the $\chi$ field. When $\mathcal{G}_{tot}$ reaches its maximum, the contribution of the $\phi$ field to the total gradient energy density is almost the same as that of the $\chi$ field, which means that the inflaton field becomes a GW source equivalent to the $\chi$ field. Thus, if the rescattering effect is not considered, the production of GWs will be seriously underestimated.

\begin{figure}
\centering
\subfigure{\label{fig9a}}{\includegraphics[width=0.48\textwidth ]{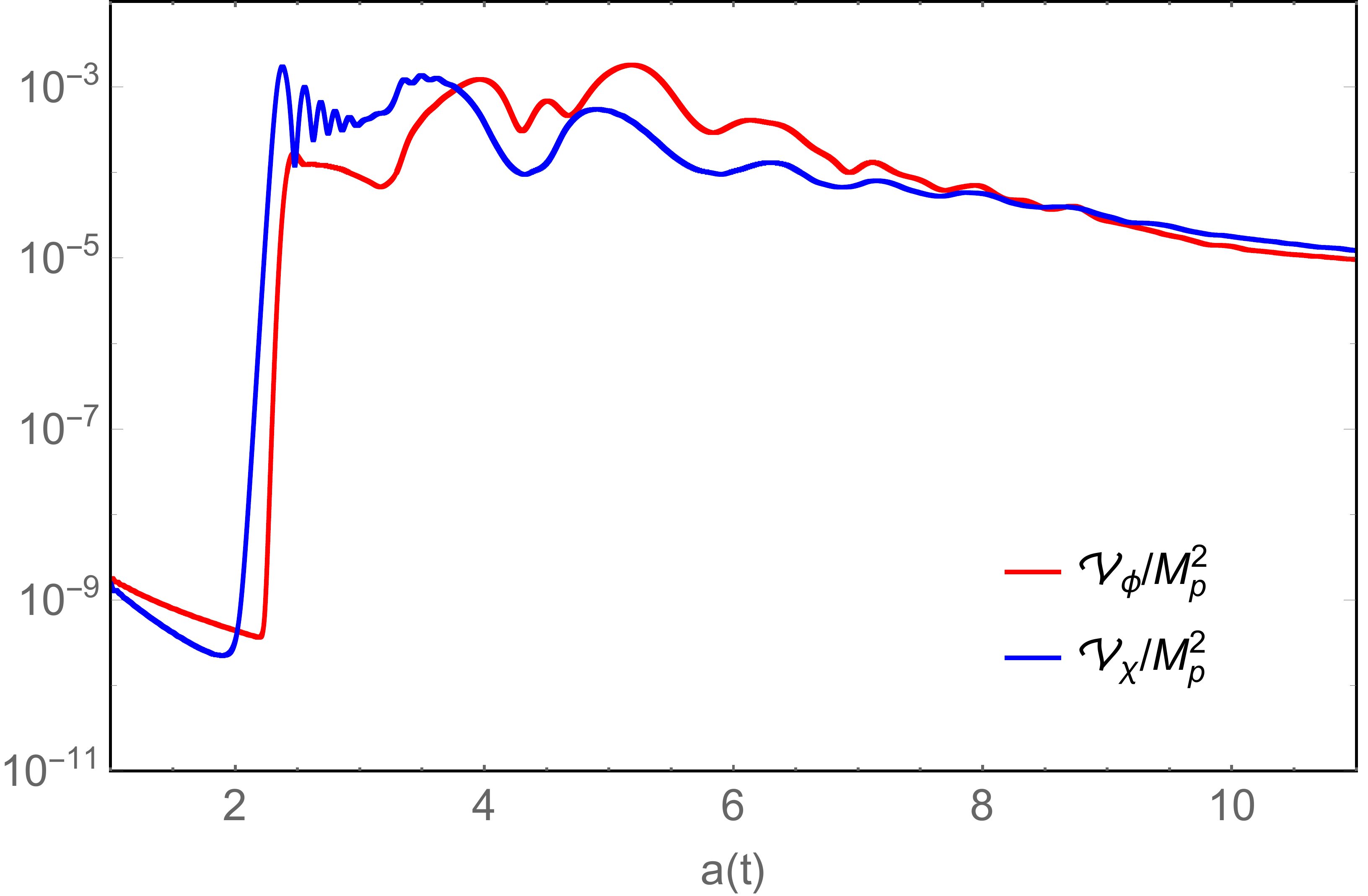}}
\subfigure{\label{fig9b}}{\includegraphics[width=0.48\textwidth ]{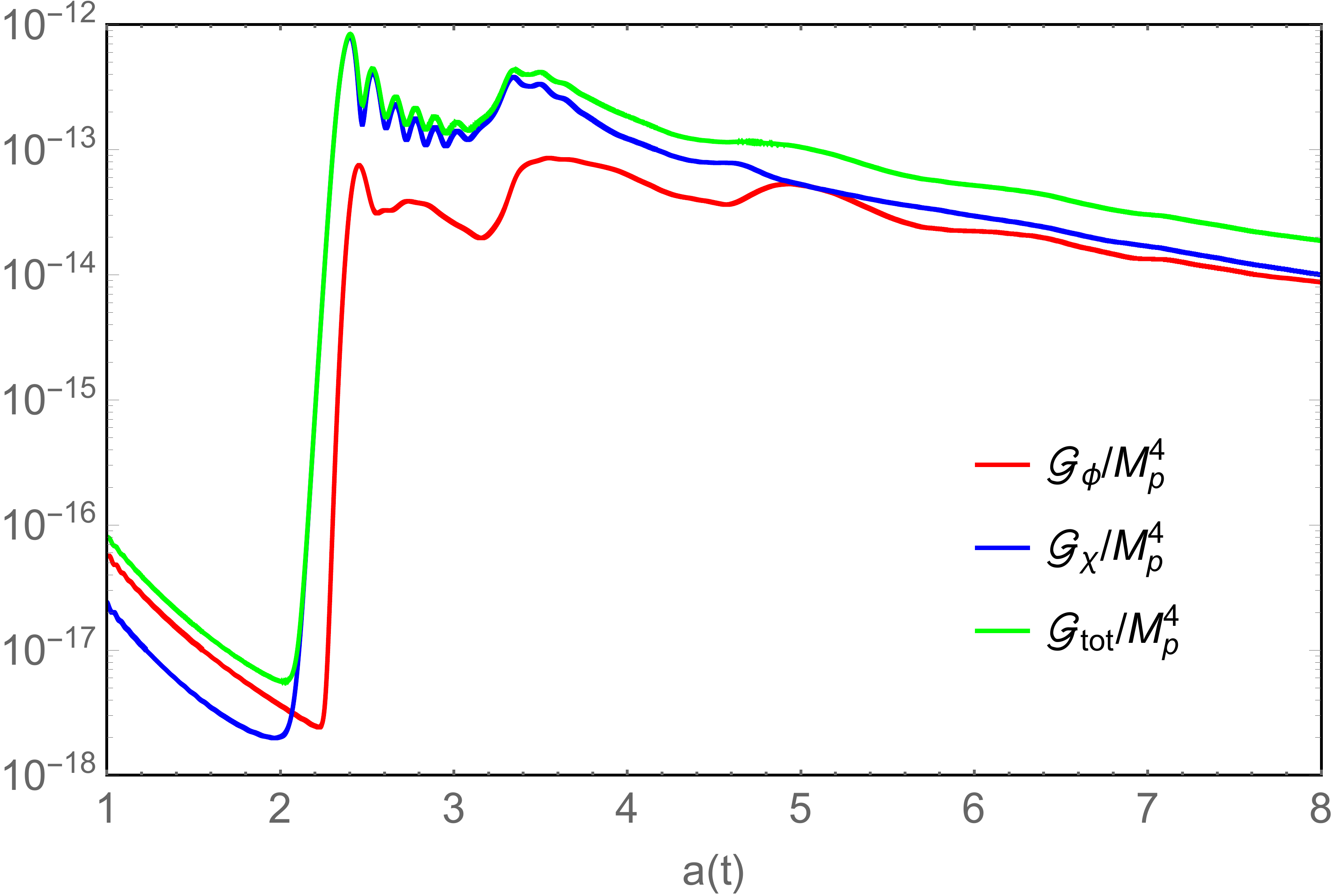}}
\caption{\label{fig9} (a) \textit{Left-hand plot}: The variances of the $\phi$ field (red) and the $\chi$ field (blue) versus $a(t)$ for $\xi=70$ and $m_\chi=0$. (b) \textit{Right-hand plot}: The evolutions of the average gradient energy density of the $\phi$ field (red line), the $\chi$ field (blue line), and their sum (green line) with  $a(t)$ for $\xi=70$ and $m_\chi=0$.}
\end{figure}

With the increase of $\xi$, since the resonance efficiency increases, more and more $\chi$ particles will be produced during each oscillation, and their backreaction will become stronger and stronger, which will decrease the number of times of exponential growth of $\mathcal{V}_\chi$. When $\xi\gtrsim65$, the variance of the $\chi$ field only needs to experience one exponential growth to reach its maximum, which can be seen in Fig. \ref{fig9a}. In this figure, the evolutions of the variances of the $\phi$ and $\chi$ fields as a function of $a(t)$ for $\xi=70$ are shown. One can see that the variance of the inflaton field is about $1$ order smaller than that of the $\chi$ field after the end of the exponential growth of $\mathcal{V}_\chi$. This is because after $\mathcal{V}_\chi$ reaches its maximum, $\mathcal{V}_\chi$ decreases quickly about $1$ order of magnitude, and thus there is not enough rescattering between the $\chi$ particles and the inflaton condensate. As a result, the $\chi$ field is the main source of GWs at the end of the exponential growth of $\mathcal{V}_\chi$. As is shown in Fig. \ref{fig9b}, when $\mathcal{G}_{tot}$ reaches its maximum, the main component of $\mathcal{G}_{tot}$ is the gradient energy of the $\chi$ field and the contribution of the inflaton field is negligible. When the Universe enters the period dominated by rescattering, the variance of the inflaton field is increased by an order of magnitude after a period of sufficient rescattering and the maximum of $\mathcal{V}_\phi$ is almost equal to that of $\mathcal{V}_\phi$ [see Fig. \ref{fig9a}]. However, from Fig. \ref{fig9b}, one can see that the gradient of the $\phi$ field during the rescattering stage does not increase significantly with respect to that at the end of the resonance. Thus, the inflaton field does not become an important source of GWs in this case, and the contribution of the inflaton field to GWs is negligible compared to the $\chi$ field. 
 
 Therefore, the inflaton field needs to meet two conditions to become an effective GW source. One is strong enough resonance, and the other is appropriate resonance efficiency.
For the positive $\xi$ case, we find that the inflaton field is a nonnegligible GW source when $6\lesssim\xi\lesssim30$.

\subsubsection{$\xi<0$}\label{3_A_2} 
\begin{figure}
\centering
\subfigure{\label{fig10a}}{\includegraphics[width=0.48\textwidth ]{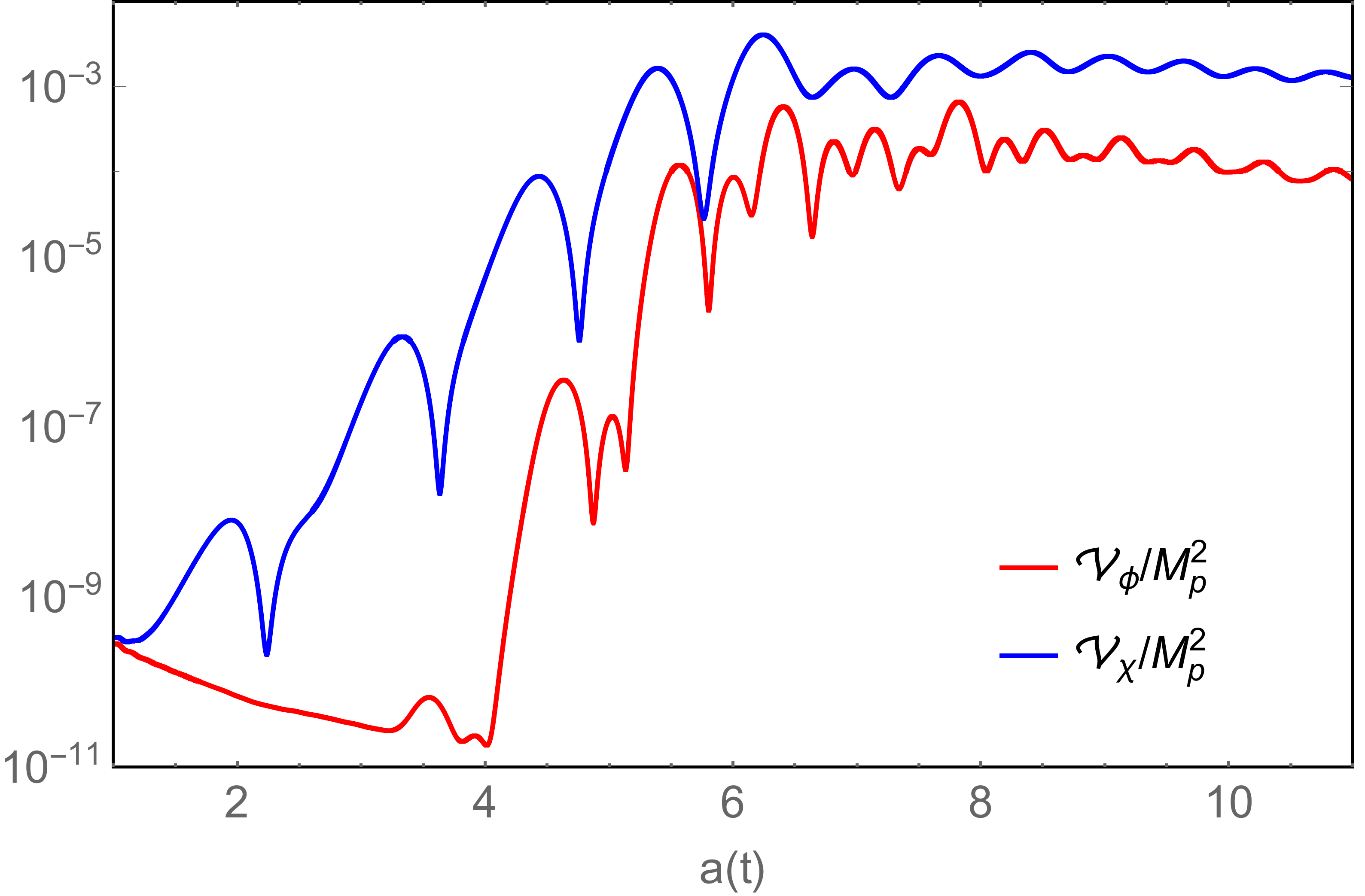}}
\subfigure{\label{fig10b}}{\includegraphics[width=0.48\textwidth ]{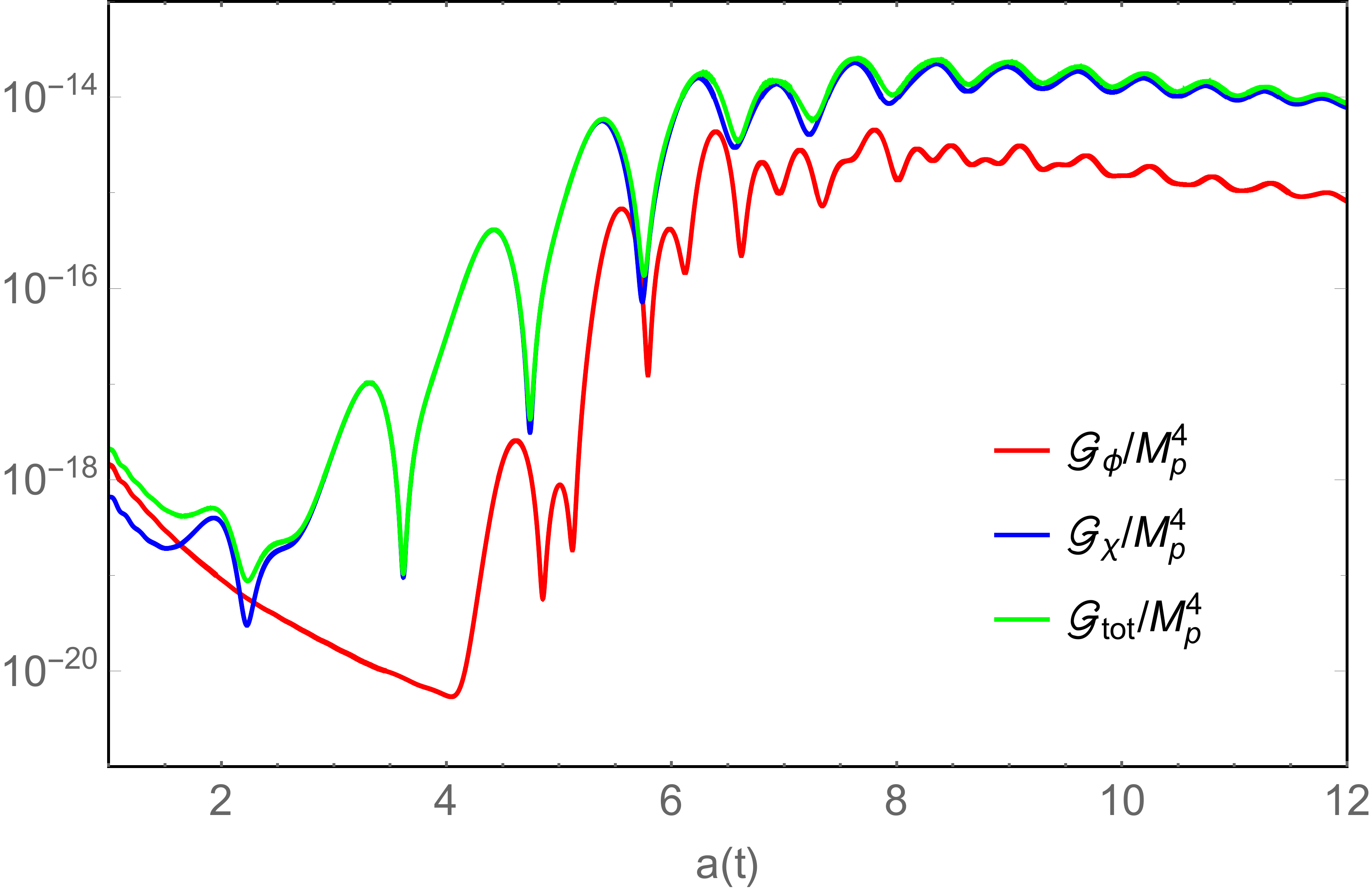}}
\caption{\label{fig10} (a) \textit{Left-hand plot}: The variances of the $\phi$ field (red) and the $\chi$ field (blue) versus $a(t)$ for $\xi=-3$ and $m_\chi=0$. (b) \textit{Right-hand plot}: The evolutions of the average gradient energy density of the $\phi$ field (red line), the $\chi$ field (blue line), and their sum (green line) with  $a(t)$ for $\xi=-3$ and $m_\chi=0$.}
\end{figure}

\begin{figure}
\centering
\subfigure{\label{fig11a}}{\includegraphics[width=0.48\textwidth ]{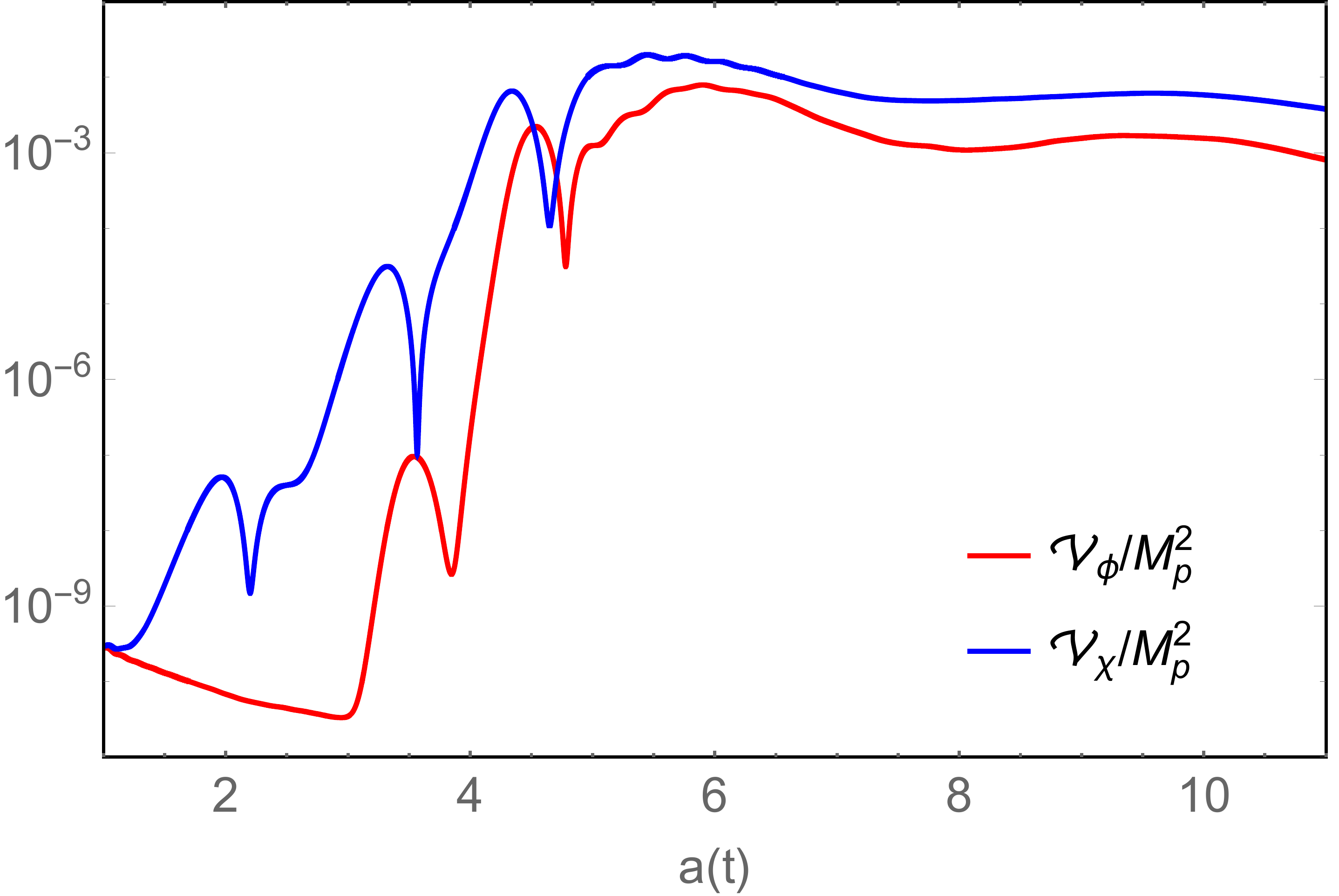}}
\subfigure{\label{fig11b}}{\includegraphics[width=0.48\textwidth ]{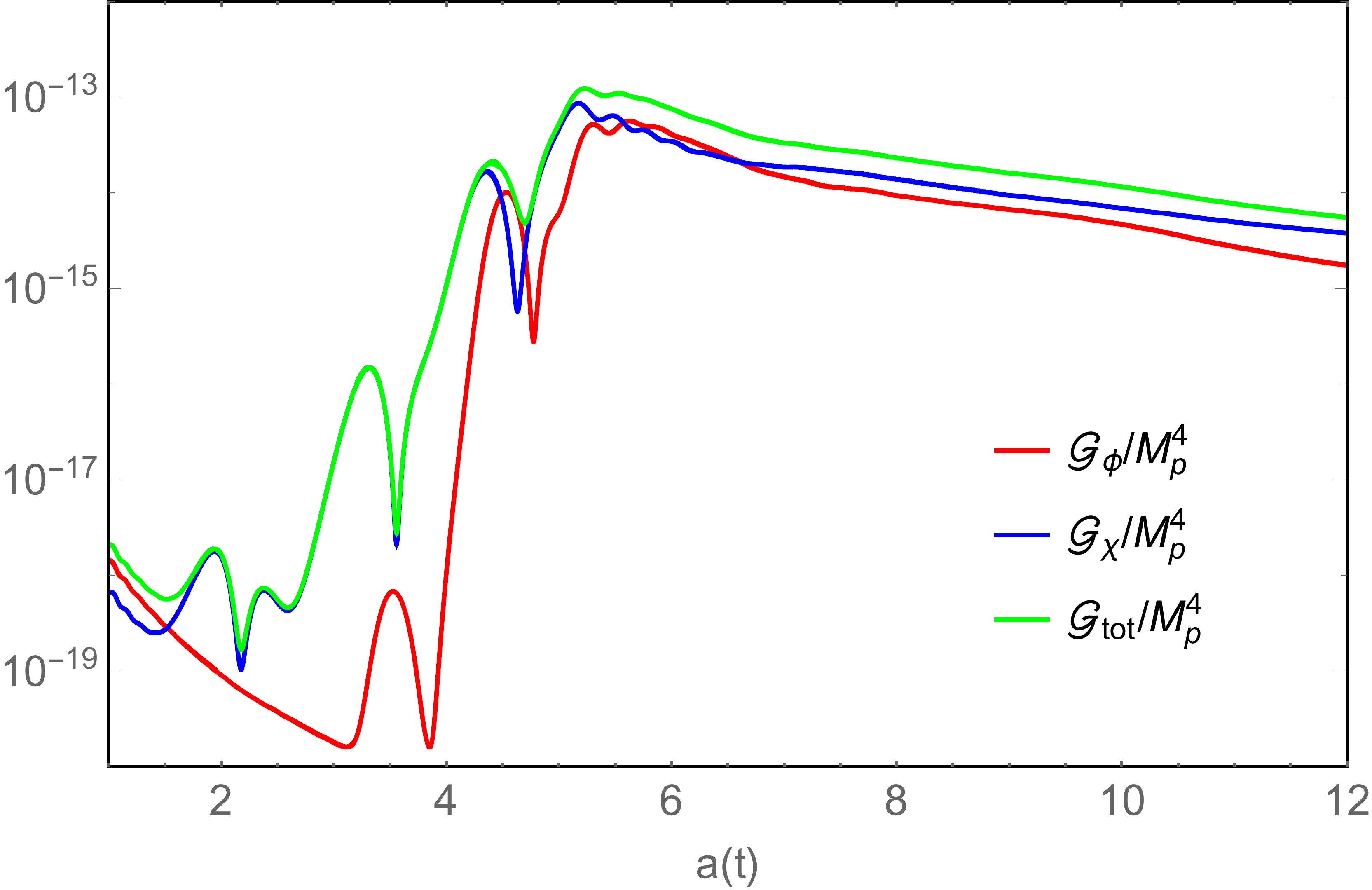}}
\caption{\label{fig11} (a) \textit{Left-hand plot}: The variances of the $\phi$ field (red) and the $\chi$ field (blue) versus $a(t)$ for $\xi=-5$ and $m_\chi=0$. (b) \textit{Right-hand plot}: The evolutions of the average gradient energy density of the $\phi$ field (red line), the $\chi$ field (blue line), and their sum (green line) with  $a(t)$ for $\xi=-5$ and $m_\chi=0$.}
\end{figure}

\begin{figure}
\centering
\subfigure{\label{fig12a}}{\includegraphics[width=0.48\textwidth ]{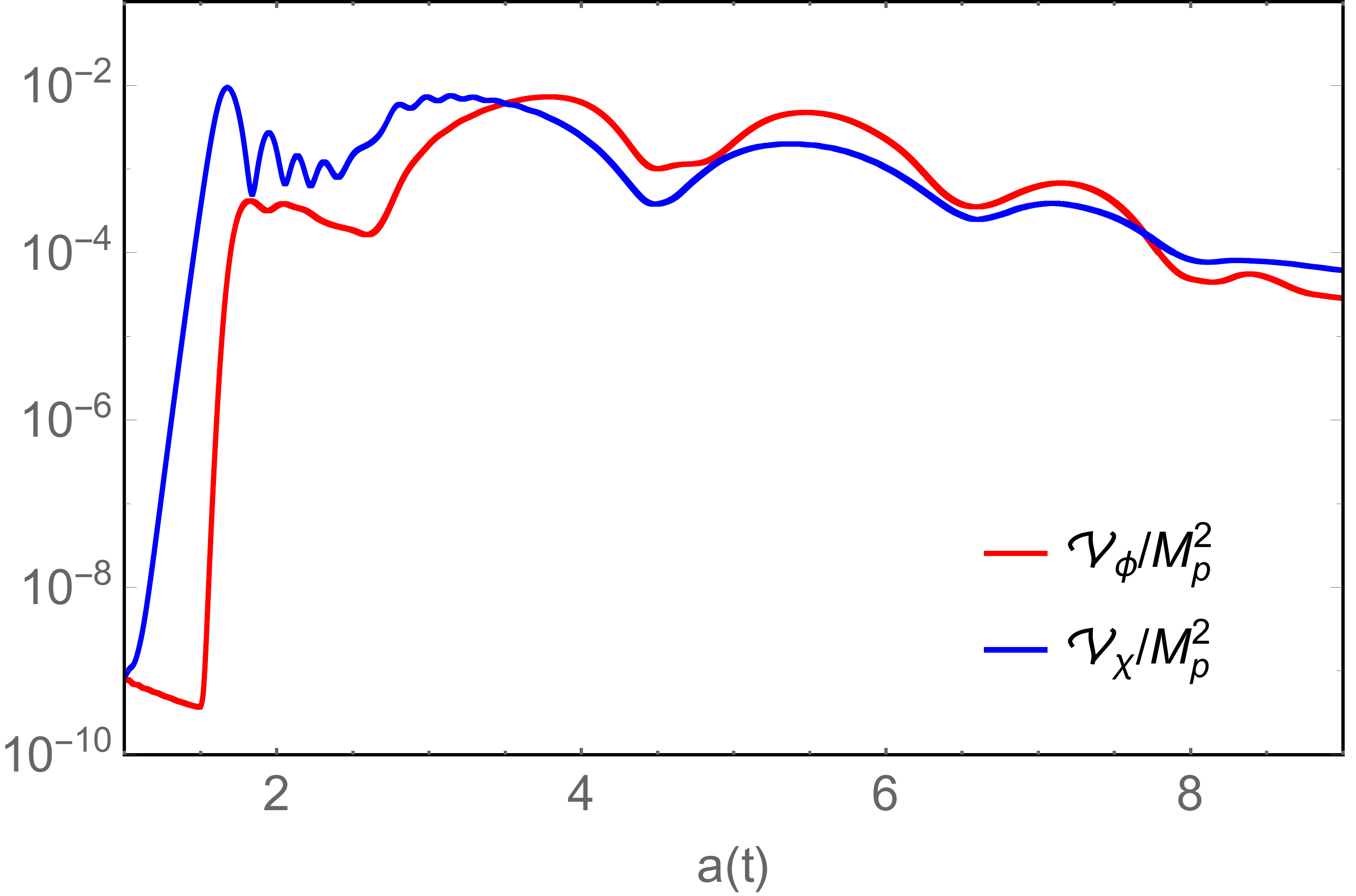}}
\subfigure{\label{fig12b}}{\includegraphics[width=0.48\textwidth ]{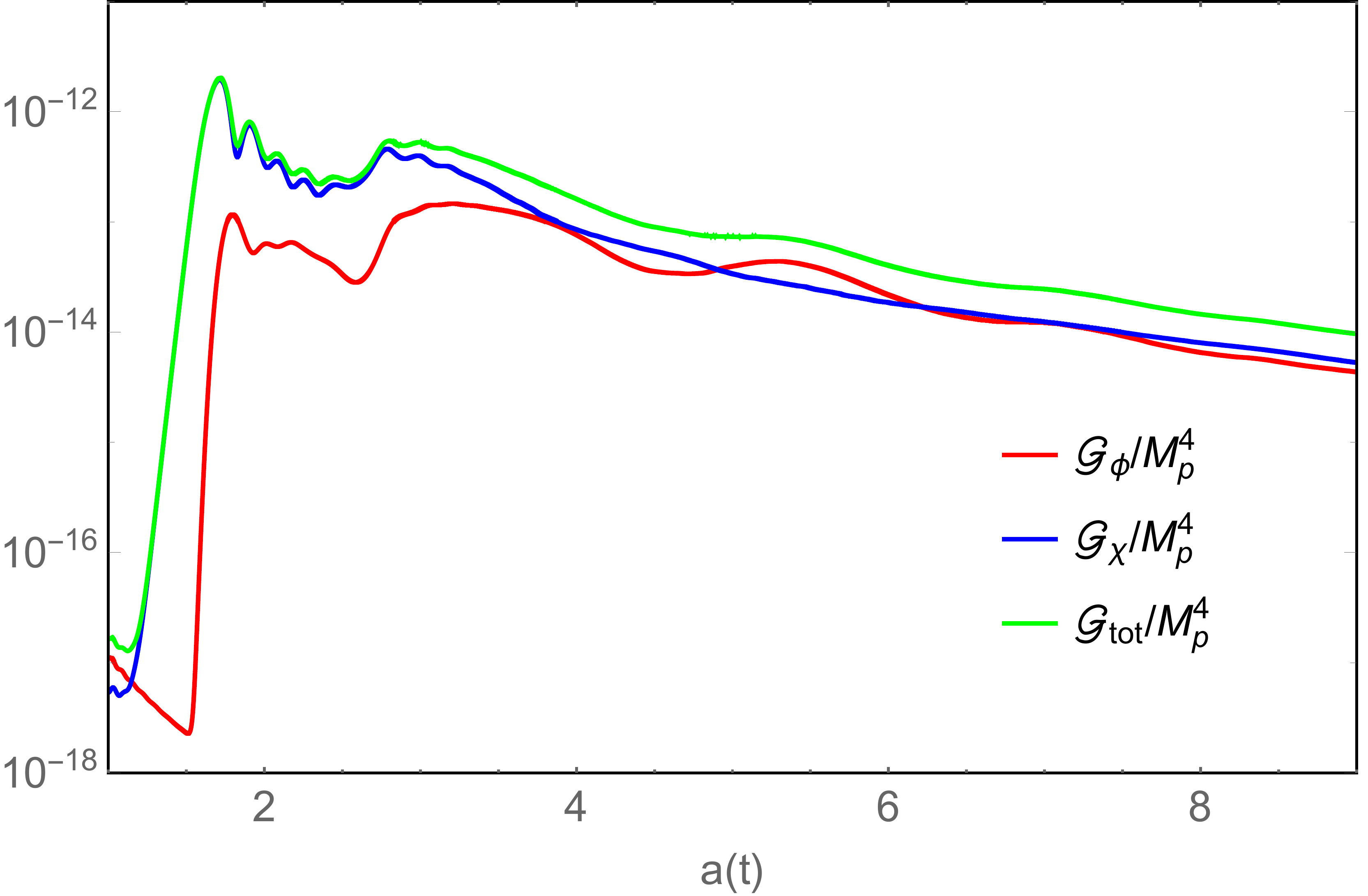}}
\caption{\label{fig12} (a) \textit{Left-hand plot}: The variances of the $\phi$ field (red) and the $\chi$ field (blue) versus $a(t)$ for $\xi=-30$ and $m_\chi=0$. (b) \textit{Right-hand plot}: The evolutions of the average gradient energy density of the $\phi$ field (red line), the $\chi$ field (blue line), and their sum (green line) with  $a(t)$ for $\xi=-30$ and $m_\chi=0$.}
\end{figure}

When $\xi<0$, all modes with $k^2/a^2<|m^2_{\chi,\text{eff}}|$ of the $\chi$ field are already tachyonic at the end of inflation, which is the main feature different from the case of the positive $\xi$ and gives the negative $\chi$ case more efficient preheating. When $\xi=-3$, the effective mass of the $\chi$ field has a  large enough amplitude to make the variance of the $\chi$ field increase before the inflaton enters the coherent oscillation stage [see Fig. \ref{fig10a}]. The resonance in the $\xi=-3$ case is far stronger than that in the $\xi=3$ case, but the contribution of the inflaton field to GWs in the $\xi=-3$ case is as negligible as that in the $\xi=3$ case [see Fig. \ref{fig10b}]. The $\xi=-3$ case is a marginal one, since when $\xi\lesssim-3$, the backreaction effect of the $\chi$ particles produced by resonance becomes dominant upon stopping  the exponential growth of $\mathcal{V}_\chi$. When $\xi=-5$, $\mathcal{V}_{\chi,\text{max}}$ is the maximum in the negative $\xi$ case, and is also larger than that obtained with $\xi=10$, which is the maximum in the positive $\xi$ case. However, unlike the $\xi=10$ case, when $\xi=-5$, the variance of the inflaton field does not increase to be equal to that of the $\chi$ field [see Fig. \ref{fig11a}]. From Fig. \ref{fig11b}, one can see that the contribution of $\mathcal{G}_\phi$ to $\mathcal{G}_{tot}$ in the $\xi=-5$ case is non-negligible but is obviously not as big as that in the $\xi=10$ case. When $\xi\lesssim-25$, the variance of the $\chi$ field  reaches its maximum after an exponential growth, which means that the growth of $\mathcal{V}_\chi$ is over before the inflaton enters the coherent oscillation stage. This property can be found in Fig.~\ref{fig12a}  where the evolutions of the variances of the $\phi$ and $\chi$ fields as a function of $a(t)$ are shown for the case of $\xi=-30$. In Fig.~\ref{fig12b}, we give the evolutions of   $\mathcal{G}_\phi$, $\mathcal{G}_\chi$ and $\mathcal{G}_{tot}$ for $\xi=-30$. One can see that, since $\mathcal{G}_\phi$ is far less than $\mathcal{G}_\chi$ after the end of the resonance and $\mathcal{G}_\phi$ does not have a significant growth during the period dominated by rescattering, the inflaton field contributes little to the GW production. For the negative $\xi$ case, we find that the inflaton field is a non-negligible GW source only when $-20\lesssim\xi\lesssim-4$.

\subsection{Massive $\chi$ particle case }
In this case, if  the backreaction of the $\chi$ field is neglected,   Eq.~(\ref{4}) can be reduced  to be
\begin{align}
m^2_{\chi,\text{eff}}\simeq e^{-2\sqrt{\frac{2}{3}}\kappa\phi}\left[3\mu^2\xi\left(e^{\sqrt{\frac{2}{3}}\kappa\phi}-1\right)+m_{\chi}^2\right]\;.
\end{align}
The tachyonic effective mass of the $\chi$ field comes from the oscillation term $(e^{\sqrt{2/3}\kappa\phi}-1)$, whose evolution as a function of $a(t)$ is plotted in Fig. \ref{13}. It is obvious that the appearance of the bare mass of the $\chi$ field will suppress the amplitude of the effective mass  and thus weaken the parametric resonance. 
  In order to achieve an effective parametric resonance, the bare mass $m_\chi$   needs to  satisfy  the following inequalities:
\begin{align}
&\left[3\mu^2\xi\left(e^{\sqrt{\frac{2}{3}}\kappa\phi}-1\right)+m_{\chi}^2\right]_{\text{min}}-\left[6\mu^2\left(e^{\sqrt{\frac{2}{3}}\kappa\phi}-1\right)\right]_{\text{min}}\lesssim 0\;\;\;(\xi>0)\;,\label{16} \\
&\left[3\mu^2\xi\left(e^{\sqrt{\frac{2}{3}}\kappa\phi}-1\right)+m_{\chi}^2\right]_{\text{min}}-\left[-3\mu^2\left(e^{\sqrt{\frac{2}{3}}\kappa\phi}-1\right)\right]_{\text{min}}\lesssim 0\;\;\;(\xi<0)\;,\label{17}
\end{align}
where the subscript ``min'' denotes the minimum. For the $\xi>0$ case, we obtain the constraint on $m_\chi$ from Eq. (\ref{16})
\begin{align}\label{18}
m_\chi & \lesssim \sqrt{3(2-\xi)\mu^2\times\left(e^{\sqrt{\frac{2}{3}}\kappa\phi}-1\right)_{\text{min}}} \nonumber\\ 
&\lesssim 0.7\mu\sqrt{\xi-2}\;.
\end{align}
When  $\xi<0$, the constraint on $m_\chi$ from Eq. (\ref{17}) becomes
\begin{align}\label{19}
m_\chi &\lesssim \sqrt{3(|\xi|-1)\mu^2\times\left(e^{\sqrt{\frac{2}{3}}\kappa\phi}-1\right)_{\text{max}}}\nonumber\\ &\lesssim 1.9\mu\sqrt{|\xi|-1}\;,
\end{align}
where the subscript ``max'' denotes the maximum. For example, when $\xi=10$ and $\xi=-10$, the occurrence of the parametric resonance requires that  the bare mass of the $\chi$ field satisfy $m_\chi\lesssim 2\mu$ and $m_\chi\lesssim 5.7\mu$, respectively. The left and right panels of Fig.~(\ref{fig14})  show the evolutions of the variance of the $\chi$ field with different values of $m_\chi$ as a function of $a(t)$   in the $\xi=10$ and $\xi=-10$ cases, respectively. As the bare mass of the $\chi$ field increases, the growth rate and the maximum of the variance of the $\chi$ field become smaller and smaller. The $\chi$ particle whose bare mass is $m_\chi=2\mu$ for $\xi=10$ and  $m_\chi=6\mu$ for $\xi=-10$ cannot be produced largely since the requirements given in Eqs.~(\ref{18}) and (\ref{19}) are not satisfied.

The small bare mass of the $\chi$ field may have an insignificant effect on the parametric resonance when $|\xi|$ is large enough,  i.e., $\xi\lesssim-25$ and $\xi\gtrsim65$, in which after an exponential growth $\mathcal{V}_\chi$ reaches its maximum.   For example, as shown in  Fig. \ref{fig15a}, when $\xi=-50$, the evolutions and maximums of $\mathcal{V}_\chi$ with $m_\chi=2\mu$ and $m_\chi=3\mu$ and $m_\chi=4\mu$ are almost the same as those in the case of the massless $\chi$ particle. This is because in the $\xi=-50$ case, the bare mass of the $\chi$ field is constrained to be $m_\chi\lesssim 13\mu$, and the small bare mass just accounts for a small proportion of the effective mass of the $\chi$ field. However, when the Universe enters the stage dominated by the rescattering,  $m_{\chi}$ affects the evolution of the field fluctuations. Figure \ref{fig15a} indicates that the abundance of the $\chi$ field variance decays faster and faster with the increase of $m_\chi$.  If $\mathcal{V}_\chi$ decays faster,  fewer inflaton particles will be knocked out.  Thus, the maximum of $\mathcal{V}_\phi$ decreases with the increase of $m_\chi$ and when $m_\chi$ is larger than a critical value, i.e., $m_\chi\gtrsim4\mu$ in the $\xi=-50$ case, the abundance of $\mathcal{V}_\phi$ does not grow during the whole rescattering era. This property can be found in Fig.~\ref{fig15b}. 
 Therefore, the mass of the $\chi$ field suppresses the abundance of $\mathcal{V}_\phi$ by weakening the parametric resonance and accelerating the decay of $\mathcal{V}_\chi$ during the rescattering period.

\begin{figure}
\centering
\includegraphics[width=0.8\textwidth ]{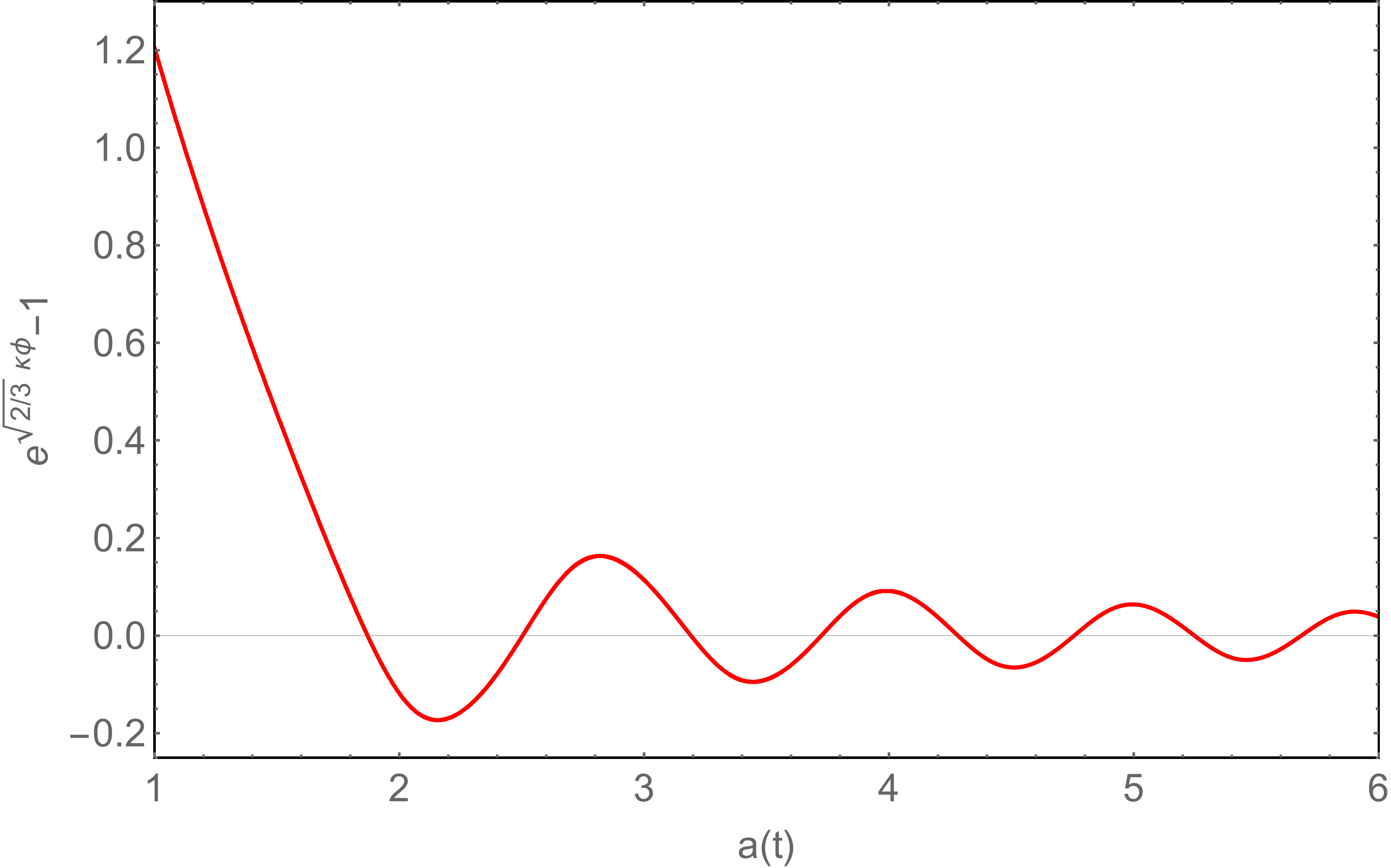}
\caption{\label{fig13} The evolution of $(e^{\sqrt{2/3}\kappa\phi}-1)$ as a function of $a(t)$.} 
\end{figure}

\begin{figure}
\centering
\subfigure{\label{fig14a}}{\includegraphics[width=0.48\textwidth ]{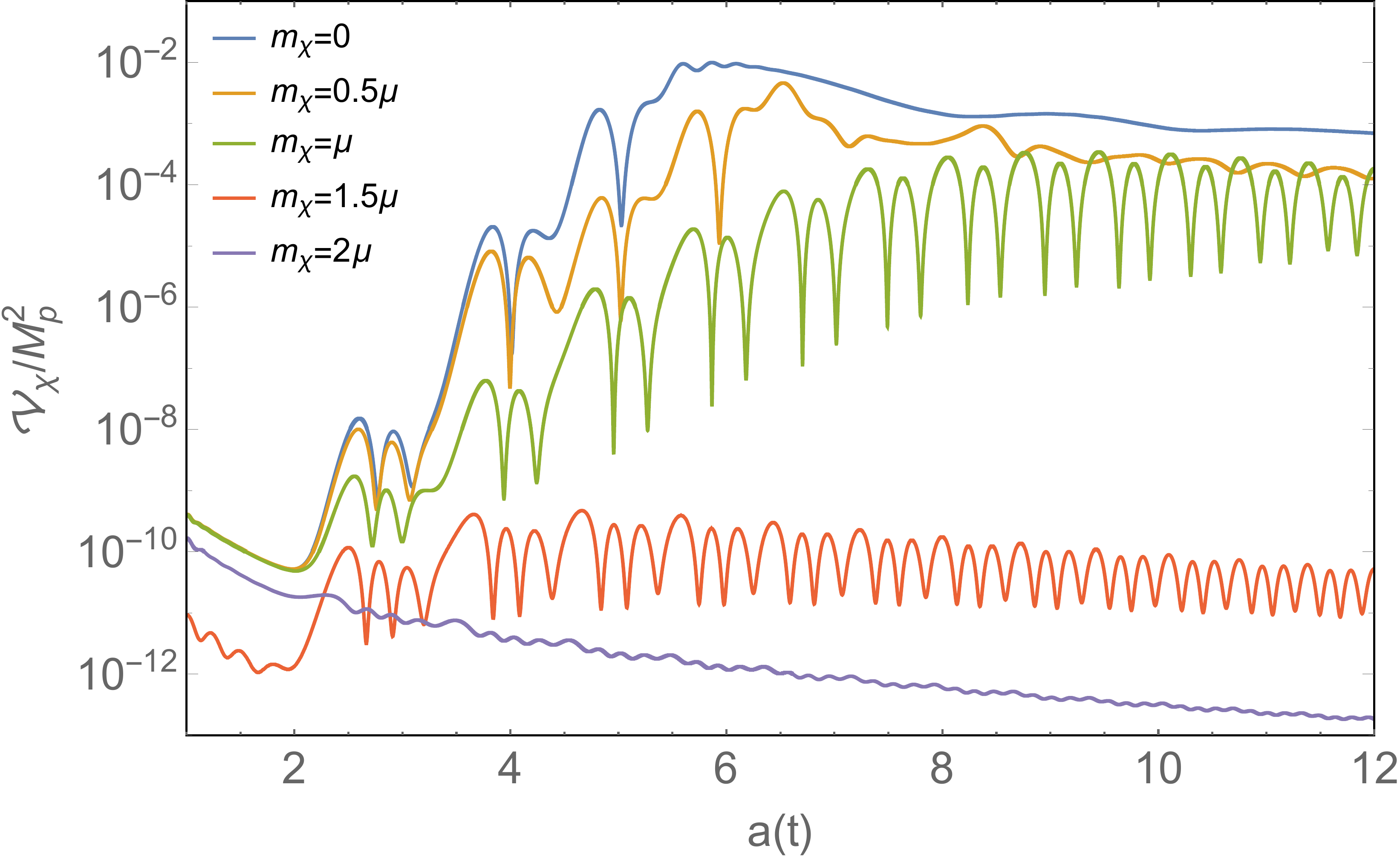}}
\subfigure{\label{fig14b}}{\includegraphics[width=0.48\textwidth ]{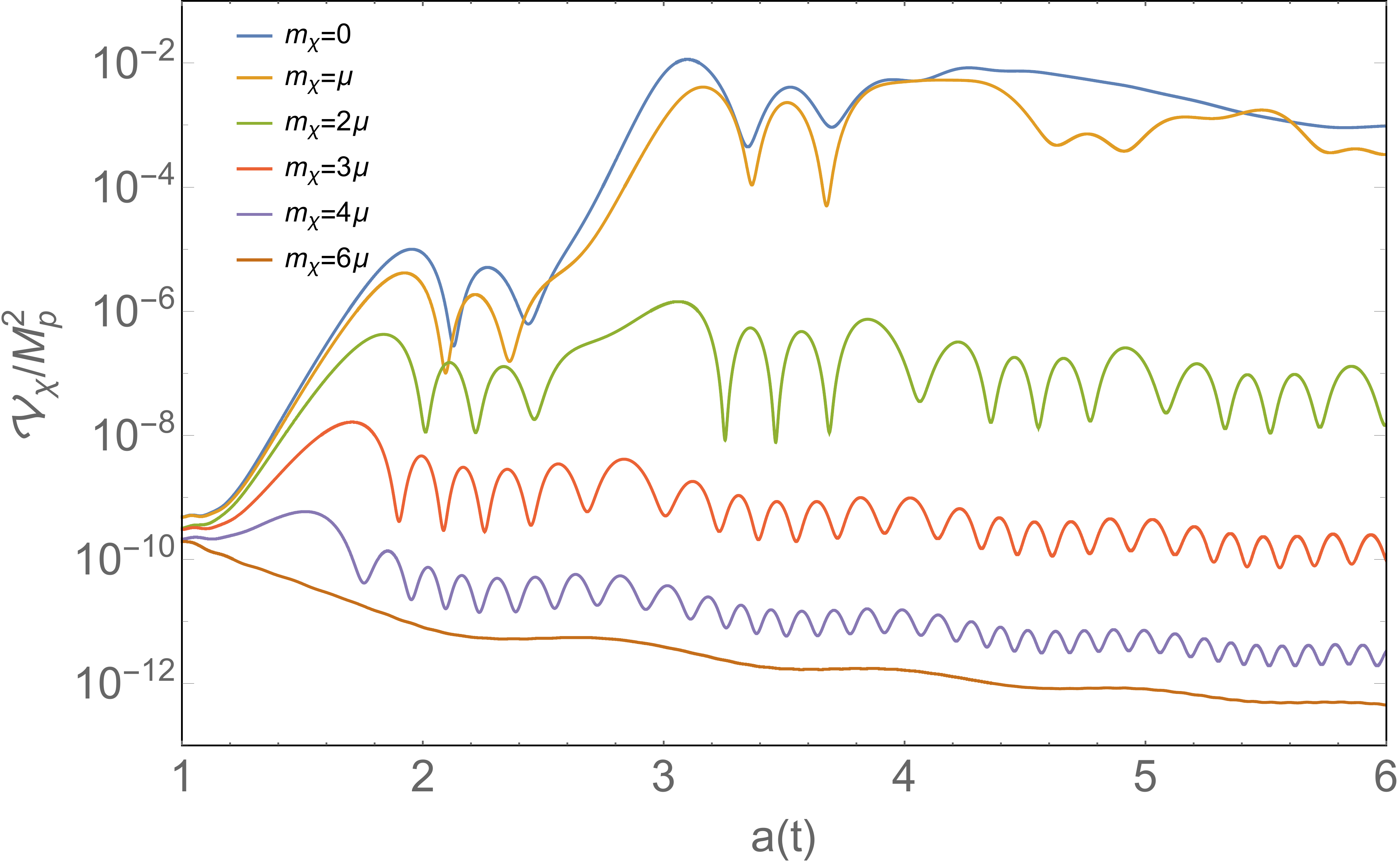}}
\caption{\label{fig14} (a) \textit{Left-hand plot}: The variances of the $\chi$ field versus $a(t)$ for $m_\chi=0$, $m_\chi=0.5\mu$, $m_\chi=\mu$, $m_\chi=1.5\mu$, and $m_\chi=2\mu$ in the $\xi=10$ case. (b) \textit{Right-hand plot}: The variances of the $\chi$ field versus $a(t)$ for $m_\chi=0$, $m_\chi=\mu$, $m_\chi=2\mu$, $m_\chi=3\mu$, $m_\chi=4\mu$, and $m_\chi=6\mu$ in the $\xi=- 10$ case.}
\end{figure}

\begin{figure}
\centering
\subfigure{\label{fig15a}}{\includegraphics[width=0.48\textwidth ]{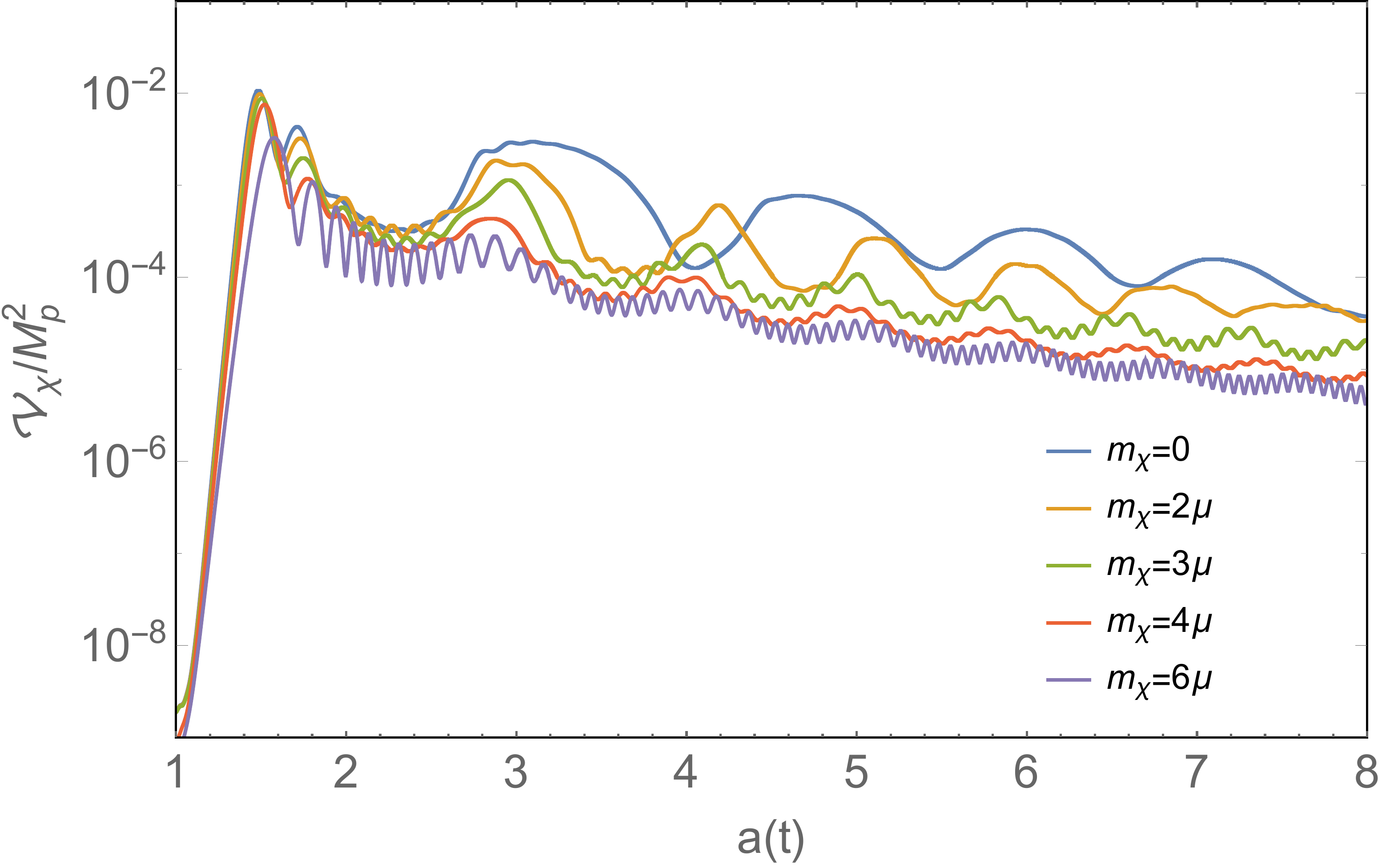}}
\subfigure{\label{fig15b}}{\includegraphics[width=0.48\textwidth ]{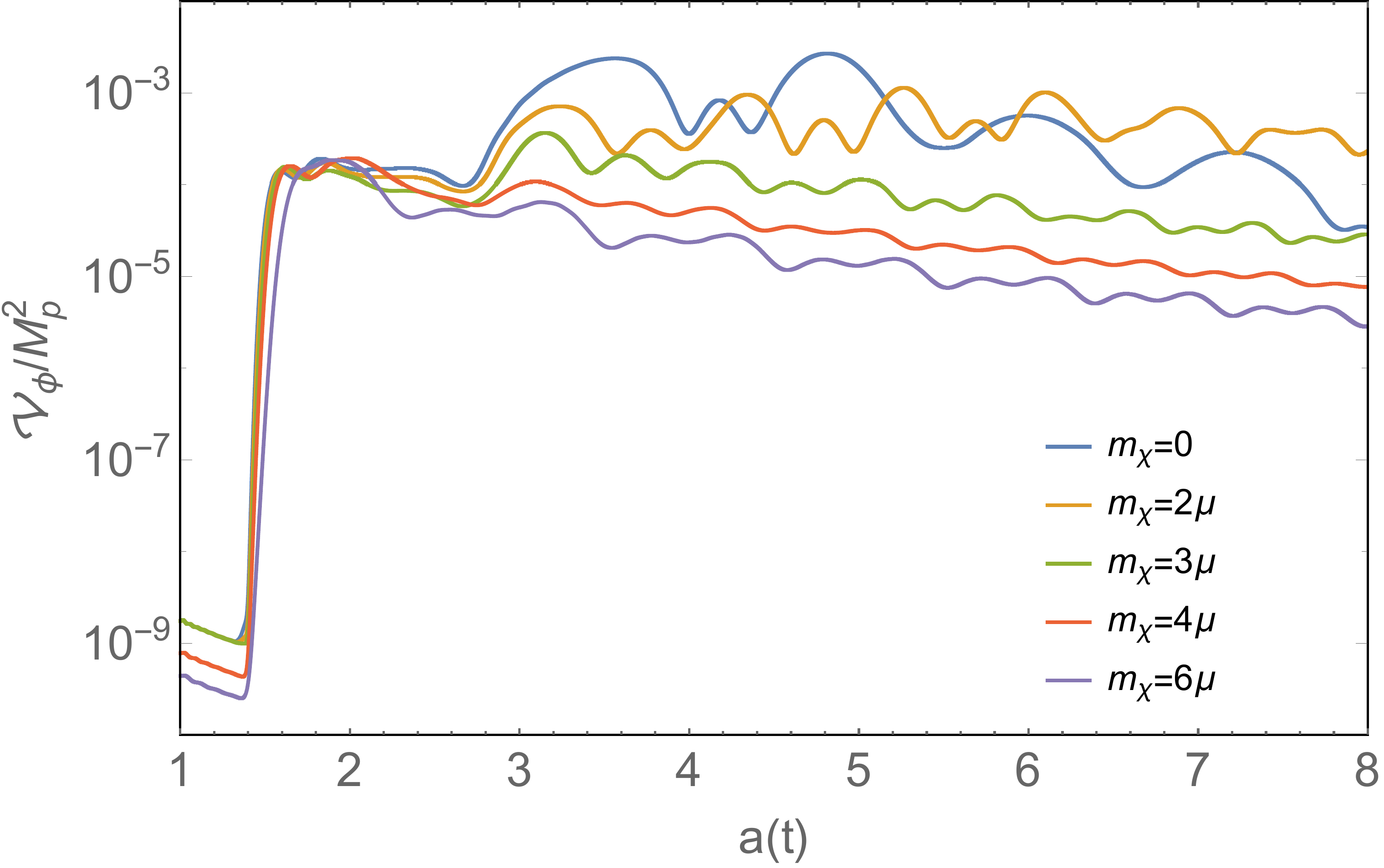}}
\caption{\label{fig15} (a) \textit{Left-hand plot}: The variances of the $\chi$ field versus $a(t)$ for $m_\chi=0$, $m_\chi=2\mu$, $m_\chi=3\mu$, $m_\chi=4\mu$, and $m_\chi=6\mu$ in the $\xi=-50$ case. (b) \textit{Right-hand plot}: The variances of the $\phi$ field versus $a(t)$ for $m_\chi=0$, $m_\chi=2\mu$, $m_\chi=3\mu$, $m_\chi=4\mu$, and $m_\chi=6\mu$ in the $\xi=-50$ case.}
\end{figure}

\section{Equation of state }\label{sec4}
Since the equation of state plays a significant role in the analysis of the matter-radiation transition during preheating, we discuss its evolution in this section. In our discussion, the spatially averaged equation of state parameter 
\begin{align}\label{20}
w\equiv\frac{\langle p \rangle}{\langle \rho \rangle}\;
\end{align}
will be analyzed.
\subsection{Massless $\chi$ particle case }
\begin{figure}
\centering
\subfigure{\label{fig16a}}{\includegraphics[width=0.48\textwidth ]{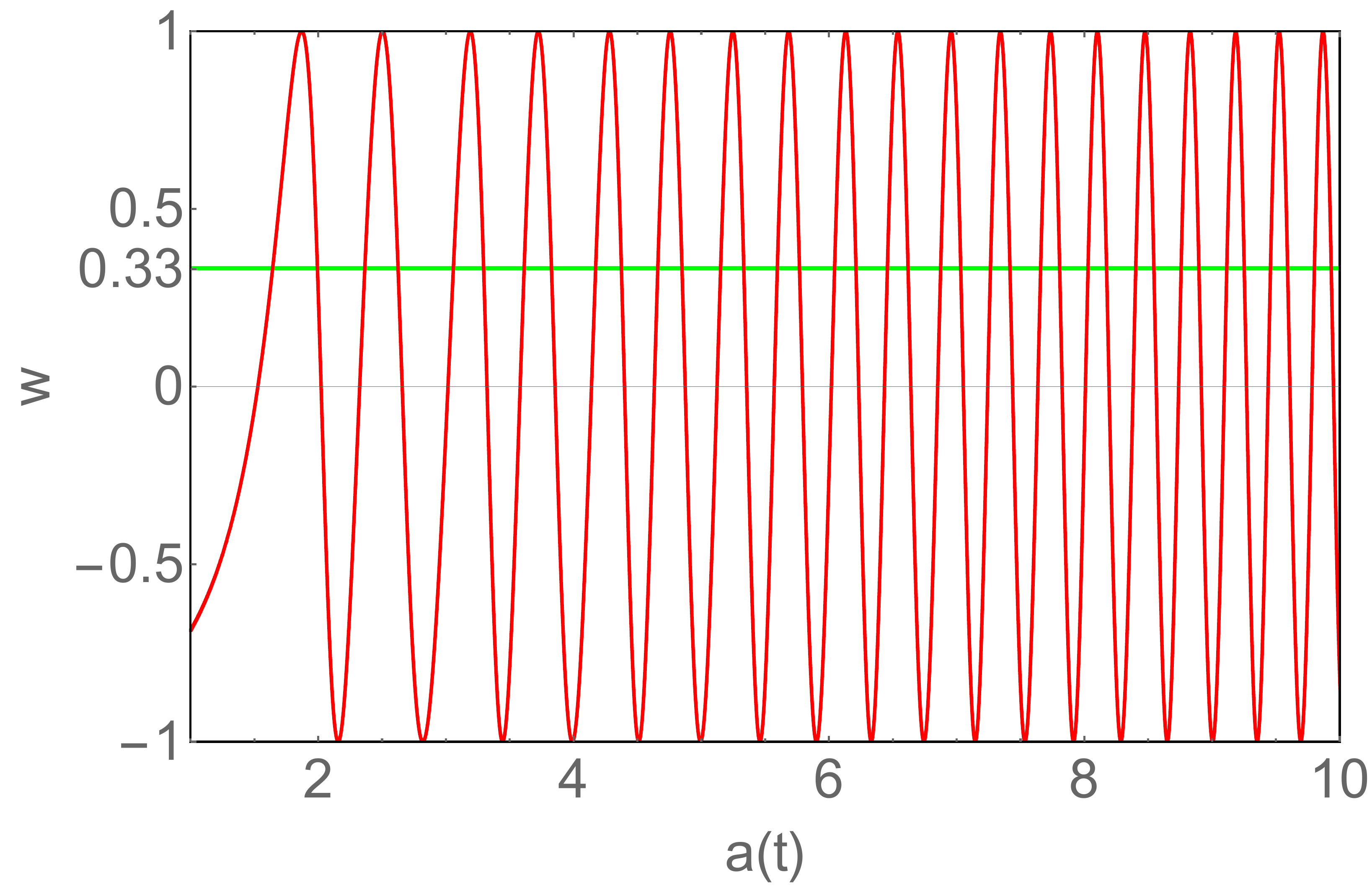}}
\subfigure{\label{fig16b}}{\includegraphics[width=0.48\textwidth ]{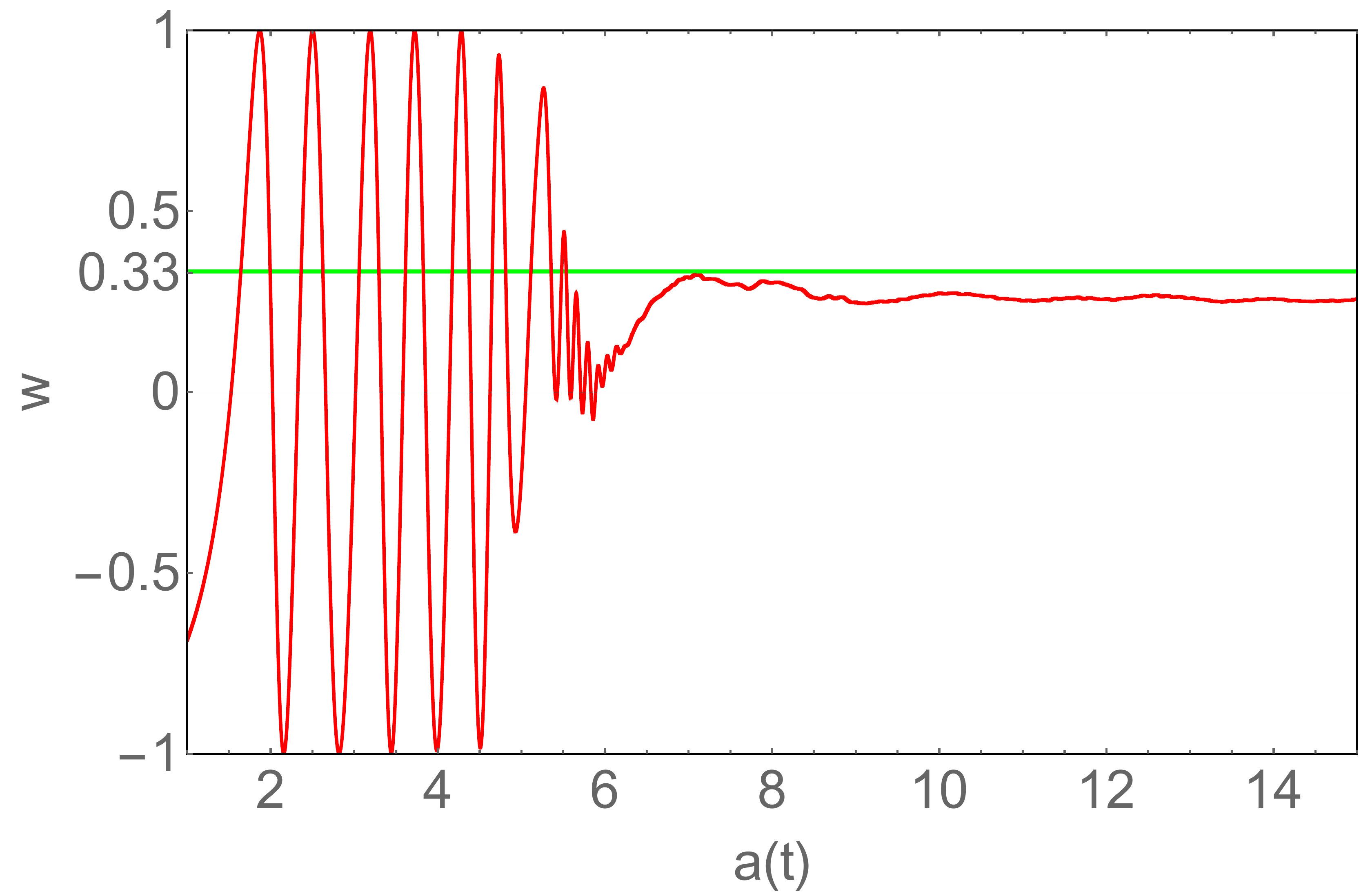}}
\caption{\label{fig16} (a) \textit{Left-hand plot}: The spatially averaged equation-of-state parameter $w$ versus $a(t)$ for $\xi=3$ and $m_\chi=0$. (b) \textit{Right-hand plot}: The spatially averaged equation-of-state parameter $w$ versus $a(t)$ for $\xi=10$ and $m_\chi=0$.}
\end{figure}

When $-3<\xi<5$, the energy is stored mainly in the homogeneous inflaton condensate due to the weak resonance, and the Universe is still dominated alternately by the potential and the kinetic energy of the inflaton field. Thus, $w$ oscillates between $-1$ and $1$, and the time average of $w$ over oscillations is zero [see Fig. \ref{fig16a}]. When $\xi\lesssim -3$ and $\xi\gtrsim 5$,  copious $\chi$ particles and inflaton particles are produced, and the Universe is no longer dominated by the inflaton condensate. Figure \ref{fig16b}, where the evolution of $w$ as a function of $a(t)$ for $\xi=10$ is plotted, indicates that the equation-of-state parameter $w$ initially oscillates between $-1$ and $1$, and eventually stabilizes at a value about $0.26$. Thus, the cosmic  phase transition occurs. 

\subsection{Massive $\chi$ particle case }
\begin{figure}
\centering
\subfigure{\label{fig17a}}{\includegraphics[width=0.48\textwidth ]{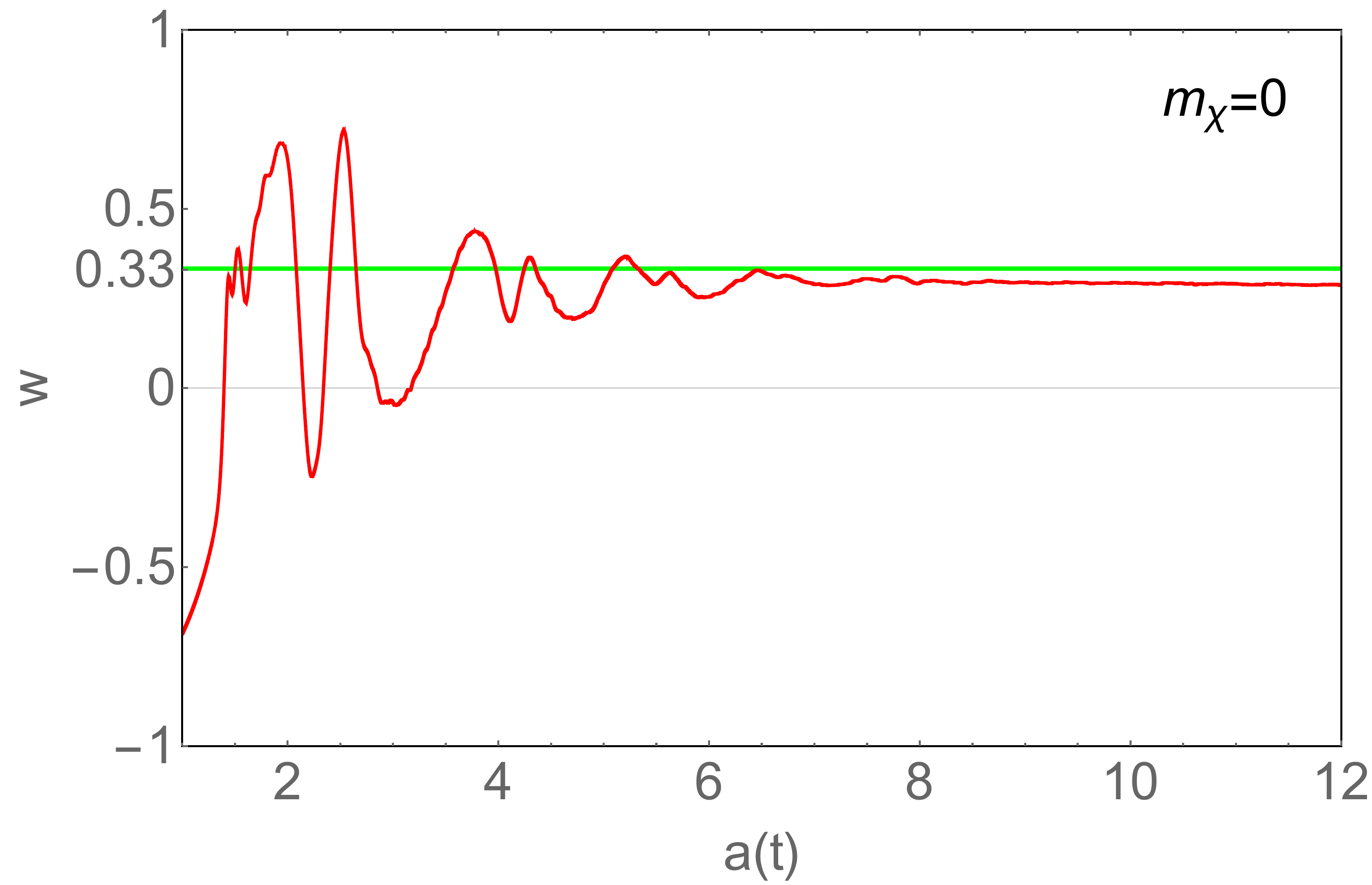}}
\subfigure{\label{fig17b}}{\includegraphics[width=0.48\textwidth ]{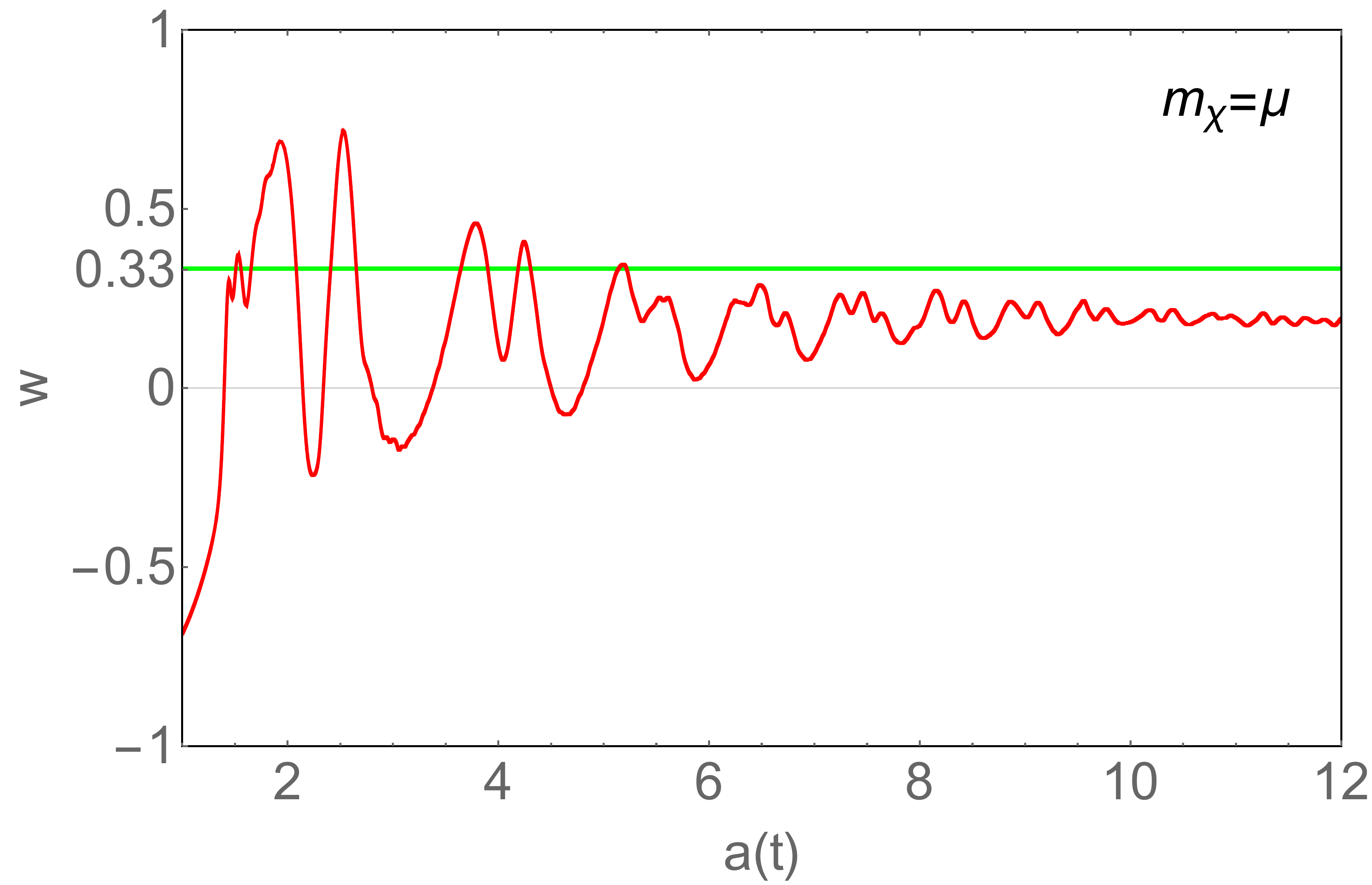}}
\subfigure{\label{fig17c}}{\includegraphics[width=0.48\textwidth ]{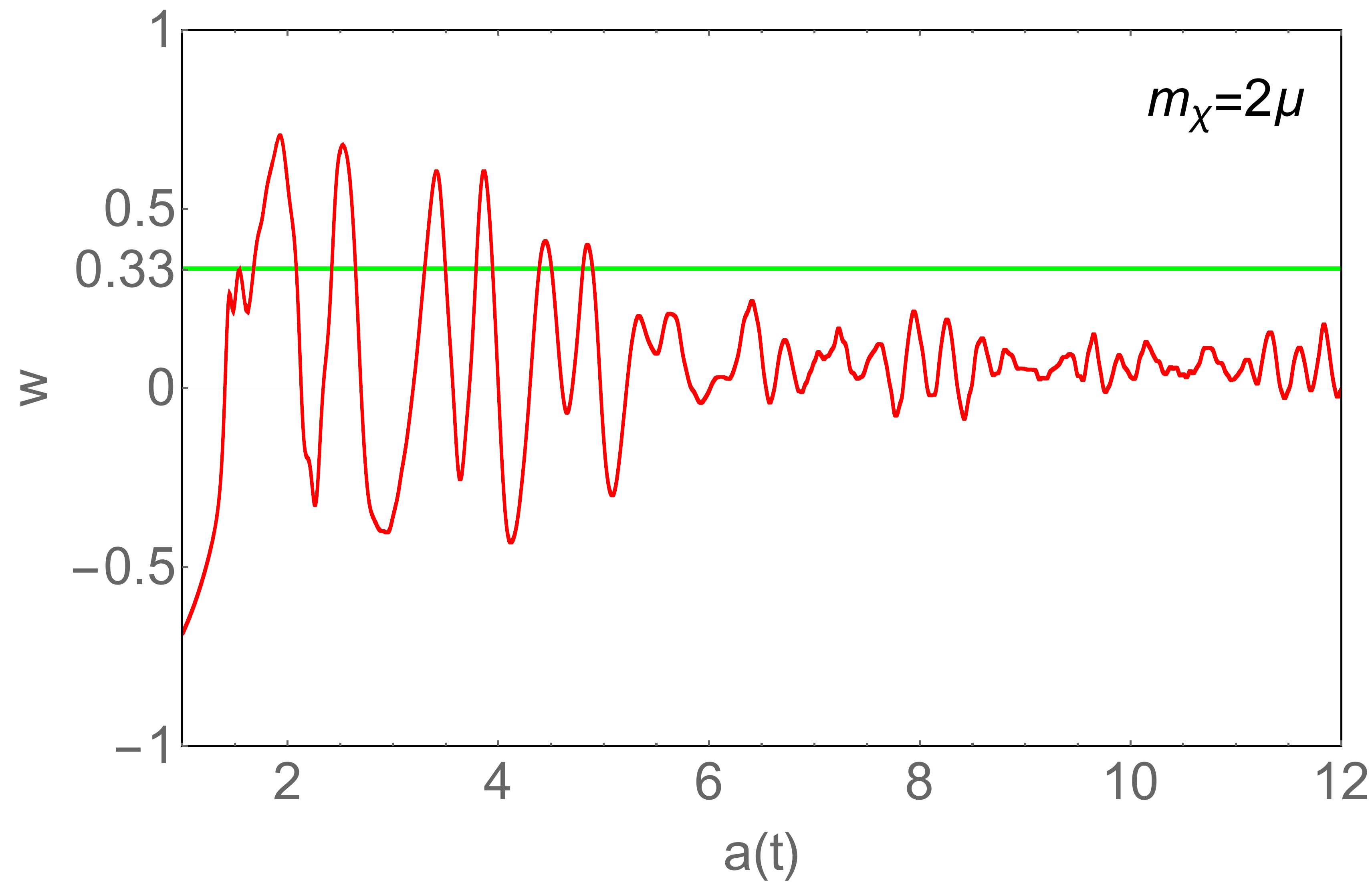}}
\subfigure{\label{fig17d}}{\includegraphics[width=0.48\textwidth ]{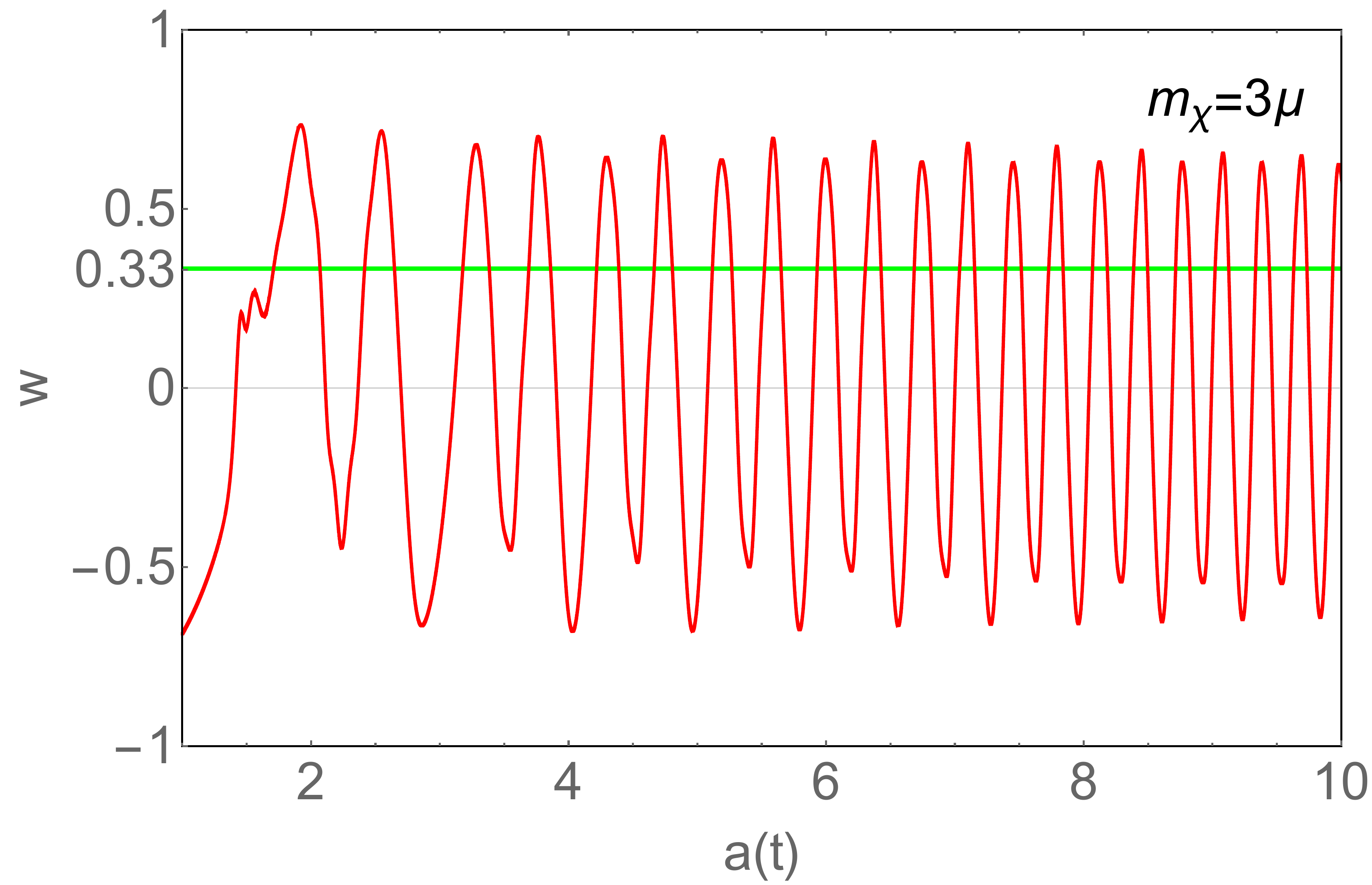}}
\caption{\label{fig17} The spatially averaged equation-of-state parameter $w$ versus $a(t)$ for $m_\chi=0$, $m_\chi=\mu$, $m_\chi=2\mu$, and $m_\chi=3\mu$ in the $\xi=-50$ case.}
\end{figure}

Figure \ref{fig17} shows the evolutions of  $w$ as a function of $a(t)$ for $m_\chi=0$, $m_\chi=\mu$, $m_\chi=2\mu$ and $m_\chi=3\mu$ in the $\xi=-50$ case. When $m_\chi=0$, $w$ will  stabilize at a value about $0.29$. For $m_\chi=\mu$, $w$ also will stabilize at a value, but this stable value is less than the one obtained in the case of $m_\chi=0$. $w$ does not completely stabilize when $m_\chi=2\mu$ and its time average  is just larger than zero. When $m_\chi=3\mu$, $w$ always keeps oscillating around zero. So, when $m_\chi=\mu$, $2\mu$, and $3\mu$, although it does not affect the parameter resonance, the nonzero mass will hinder the transition from the matter-dominated era to the radiation-dominated one. This is because $m_\chi$  suppresses the abundance of $\mathcal{V}_\phi$, which leads to  the inflaton condensate oscillating when the mass is large enough. 

Figures~\ref{fig16} and \ref{fig17} indicate clearly that the end of preheating cannot  connect smoothly with the radiation phase. This is because  reheating never completes at the phase of preheating, which is only the first stage of reheating. The $\chi$ particles produced in preheating will  decay   to the elementary particles populating and thermalizing the Universe.

\section{Evolution of the scalar metric variable}\label{sec5}
\begin{figure}
\centering
\includegraphics[width=0.8\textwidth ]{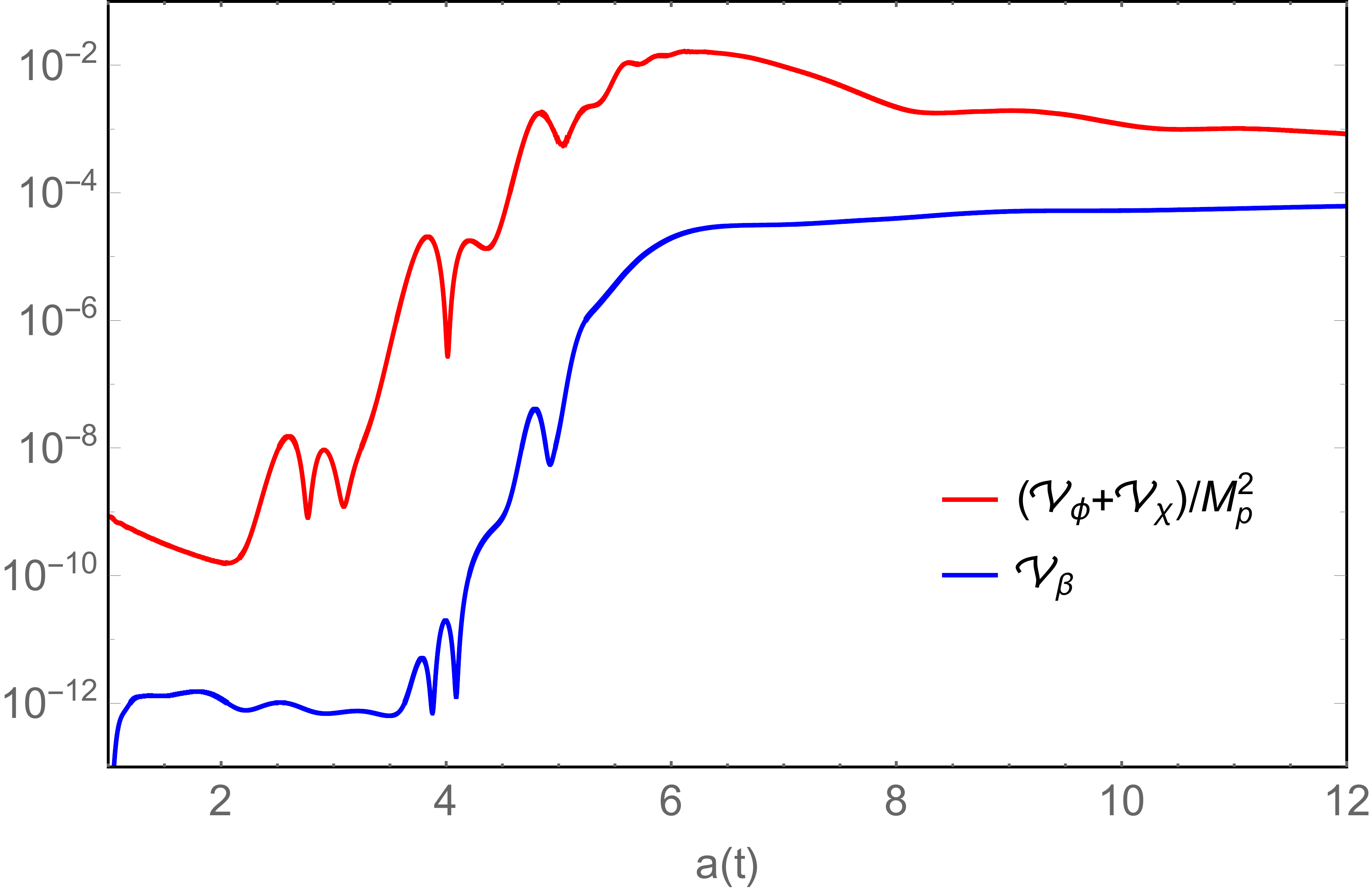}
\caption{\label{fig18} The total variance of the scalar fields (red line) and the variance of the scalar metric perturbations (blue line) versus $a(t)$ for $\xi=10$ and $m_\chi=0$.} 
\end{figure}

\begin{figure}
\centering
\includegraphics[width=0.8\textwidth ]{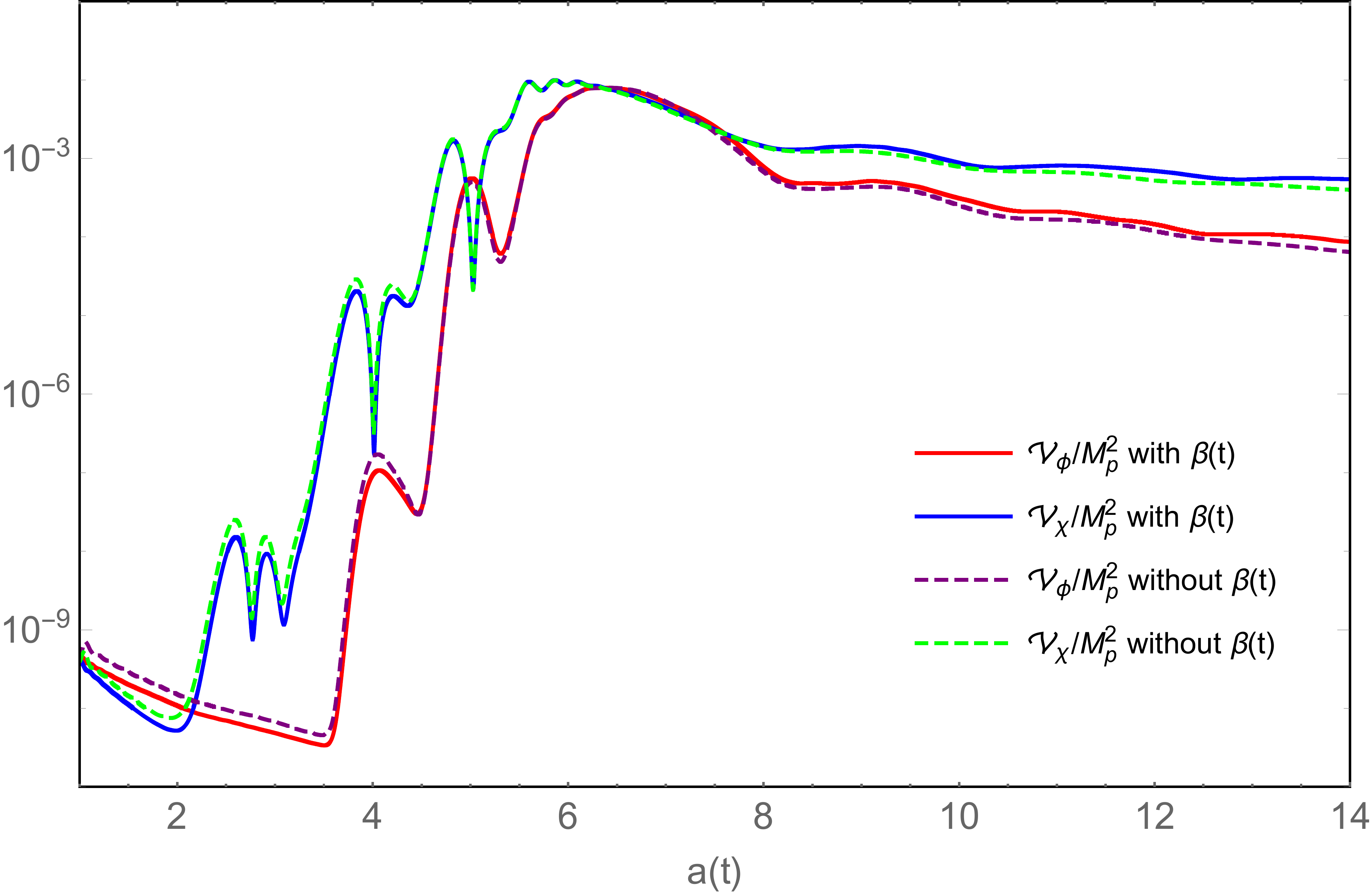}
\caption{\label{fig19} The variances of the inflaton field (solid red line) and $\chi$ field (solid blue line) versus $a(t)$ for $\xi=10$ and $m_\chi=0$. The variances of the inflaton field (dashed purple line) and $\chi$ field (dashed green line) in a FRW metric are also shown for comparison.} 
\end{figure}

Now, we investigate the effect of scalar metric perturbations. Figure \ref{fig18} shows the evolutions of  $\mathcal{V_\phi}+\mathcal{V}_\chi$ and the variance of the scalar metric fluctuations $\mathcal{V}_\beta\equiv \langle\beta^2\rangle-\langle\beta\rangle^2$ as a function of $a(t)$ for $\xi=10$ and $m_\chi=0$. It is easy to see that, although the scalar metric variable is initialized to be homogeneous, $\mathcal{V}_\beta$ quickly increases from $0$ to $10^{-12}$ after the system begins to evolve. Then the variance of $\beta$ increases exponentially with the fast growth of the total variance of the scalar fields. When the total variance stops the growth, $\mathcal{V}_\beta$ reaches a stable value. To reveal the influence of the metric fluctuations on the evolutions of the scalar fields, we plot the $a(t)$-dependent variances of the $\phi$ and $\chi$ fields with and without the scalar metric fluctuations in Fig.~\ref{fig19}.  We find that the evolutions of the scalar fields almost do not feel the appearance of the scalar metric perturbations. 

\begin{figure}
\centering
\includegraphics[width=0.8\textwidth ]{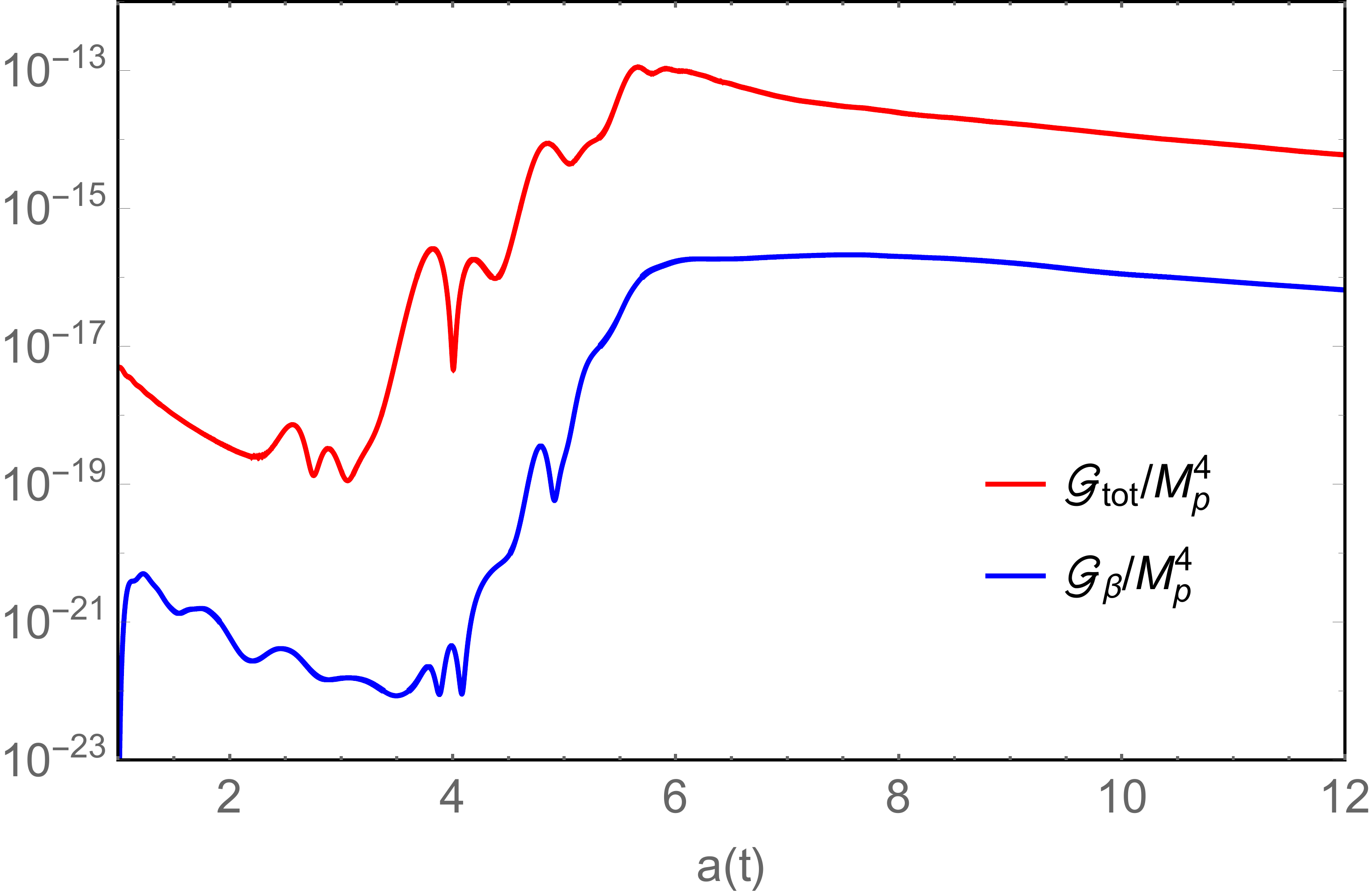}
\caption{\label{fig20} The average total gradient energy density of the scalar fields $\mathcal{G}_{tot}$ (red line) and the average gradient contribution of the scalar metric fluctuations $\mathcal{G}_{\beta}$ (blue line) versus $a(t)$ for $\xi=10$ and $m_\chi=0$.} 
\end{figure}

The large enhancement of the scalar metric perturbations will induce the production of the second-order GWs, and the contribution to GWs of the scalar metric perturbations is associated with their gradient~\cite{2008Bastero-Gil,2010Bastero-Gil}. If the gradient of the scalar metric perturbations can be comparable to that of the scalar fields,  they become a non-negligible  GW source.
In Fig. \ref{fig20}, we plot the evolutions of the average total gradient energy density of the scalar fields $\mathcal{G}_{tot}$ and the average gradient contribution of the scalar metric fluctuations $\mathcal{G}_{\beta}\equiv \langle e^{-2\beta}(\partial \beta)^2\rangle/(2M_p^{-2})$ as a function of $a(t)$ for $\xi=10$ and $m_\chi=0$. Since the gradient of $\beta$ is far less than that of the scalar fields, the scalar metric perturbations cannot become a significant GW source. Similar results are found in all other cases.

Since the scalar metric perturbations have no effect on the evolutions of the scalar fields and the production of GW, it is reasonable to conjecture that the contributions of other kinds of metric perturbations can also be neglected  during preheating. Thus, one can investigate the spectrum of GWs with a simple FRW metric during preheating in this model, which is an interesting issue but is beyond the scope of the present work.

\section{conclusions and discussions}
\label{sec6}
The Starobinsky model is now strongly favored among the models of inflation. However, after the Universe enters into the reheating period, the parametric resonance is absent in a pure Starobinsky model. In Ref. \cite{1999Tsujikawa1}, a new scalar field $\chi$ coupled to the curvature scalar is introduced and the linear preheating is investigated by using the Hartree approximation. To consider the effects of the rescattering and the metric perturbations neglected in Ref. \cite{1999Tsujikawa1}, in this paper, we have investigated the preheating with a three-dimensional lattice simulation.  
 We find that the rescattering between the produced $\chi$ particles and the inflaton condensate makes the maximum of the $\chi$ field variance bigger than that in the Hartree approximation. Meanwhile, the copious inflaton particles can be knocked out of the inflaton condensate by rescattering. Thus, the inhomogeneous inflaton field may become  a significant GW source. When the scalar field $\chi$ is massless, the contribution of the inflaton field to GWs is non-negligible compared with that of the $\chi$ field when $6\lesssim\xi\lesssim 30$ and $-20\lesssim\xi\lesssim -4$. For the massive $\chi$ particle case, the nonzero mass will suppress the parametric resonance, and accelerate the decay of the variance of the $\chi$ field during the rescattering period. These effects suppress the abundance of the variance of the inflaton field and hinder the occurrence of the cosmic phase transition from matter to radiation. Finally, we find that the sub-Hubble scalar metric fluctuations do not affect the evolutions of the scalar fields, and cannot become an effective GW source for the model considered in this paper.

Finally,  we must point out that  only  the parametric resonance of the $\chi$ field fluctuations on subhorizon scales is considered in the present paper. Indeed, the super-Hubble modes of the $\chi$ field will also be amplified due to the tachyonic instability, which may lead to an amplification of super-Hubble scalar metric fluctuations. Recently, Ref. \cite{2018cai} has shown, by the numerical method, that the super-Hubble metric perturbations of scalar type enhanced during the preheating with  an arbitrary power-law potential could significantly affect the theoretical predictions for the CMB observations. 
 And a similar result was earlier derived semianalytically in Ref. \cite{2014cai} by virtue of the covariant method. Therefore, for the model considered in the present paper, the evolution of the super-Hubble scalar metric perturbations during preheating is expected to have the same effect, which is an interesting issue  left for future work.

\begin{acknowledgments}
We appreciate very much the insightful comments and helpful suggestions by an anonymous referee, and thank Professor Shuang-Yong Zhou very much for fruitful discussions. This work was supported by the National Natural Science Foundation of China under Grants No. 11775077, No. 11435006, and No. 11690034.
\end{acknowledgments}

\end{document}